\documentclass[fleqn,usenatbib]{mnras}

\usepackage{graphicx}
\usepackage{amsmath}
\usepackage{amssymb}
\usepackage{amsfonts}
\usepackage{mathrsfs}
\usepackage{bm}
\usepackage{booktabs}
\usepackage{hyperref}
\usepackage{url}
\usepackage{xcolor}

\hypersetup{colorlinks=true,citecolor=blue,urlcolor=blue,linkcolor=blue}

\providecommand{\ion}[2]{#1\,{\scshape #2}}

\title[Generalised Polytropic and Warm Dark Matter in NGC~5055]
      {Studying Generalised Polytropic and Warm Dark Matter in NGC~5055}

\author[A. Sanyal, M. K. Hossain and F. Rahaman]
{Aritra Sanyal$^{1}$\thanks{E-mail: aritrasanyal1@gmail.com},
Md Khalid Hossain$^{1}$\thanks{E-mail: mdkhalid.ac@gmail.com},
Farook Rahaman$^{1}$\thanks{E-mail: rahaman@associates.iucaa.in}
\\
$^{1}$Department of Mathematics, Jadavpur University, Kolkata 700032, India
}

\date{Accepted XXX. Received YYY; in original form ZZZ}
\pubyear{2026}

\begin{document}
\label{firstpage}
\pagerange{\pageref{firstpage}--\pageref{lastpage}}
\maketitle

\begin{abstract}
Galaxy rotation curves provide a powerful probe of mass distributions in spiral
galaxies. We present a general relativistic analysis of NGC~5055 using its observed
rotation curve and WISE W1 ($3.4\,\mu$m) photometry to constrain the stellar mass
profile. Adopting a static, spherically symmetric spacetime with anisotropic matter
(vanishing radial pressure), we fit a modified exponential azimuthal velocity law to
kinematic data, reconstructing metric functions and deriving enclosed mass, energy
density, and tangential pressure profiles. The stellar mass underpredicts the total
gravitating mass at intermediate-to-large radii, indicating dominant dark matter.
The physical viability of the resulting relativistic model is examined through energy
conditions, causality constraints, and a generalised polytropic equation of state
analysis. Among the candidate equations of state tested --- Bose--Einstein condensate,
Chaplygin gas, linear fluid, and generalised polytropic --- only the generalised
polytropic form $p = K\rho^{\gamma}$ with $\gamma = 0.91$ and
$K = 4.63\times10^{-8}$ achieves an excellent fit , firmly
placing the dark matter of NGC~5055 in the Warm Dark Matter regime. All energy
conditions and causality constraints are satisfied, and the halo is stable against
gravitational cracking. Sterile neutrinos or gravitinos of keV-scale mass are
identified as the most physically motivated particle candidates. Overall, this study
demonstrates that combining rotation-curve data with photometric stellar mass
estimates within a general relativistic framework provides a consistent and physically
viable description of the mass distribution in spiral galaxies such as NGC~5055.
\end{abstract}


\section{Introduction}
\label{sec:intro}

Galaxies are gravitationally bound systems of stars, gas, dust, and dark matter.
Observationally, matter is not homogeneously distributed throughout the Universe but
instead remains confined within self-contained structures spanning a wide range of
morphologies and spatial scales. Based on their global properties, galaxies are
commonly classified as spiral, elliptical, lenticular, irregular, or starburst
systems. Understanding the physical mechanisms that maintain these bound structures
is a central problem in galactic astrophysics and is closely related to the
distribution of dark matter within galactic halos
\citep{Rubin_1980,Sofue_Rubin_2001,deBlok_2010}.

General relativity provides a fundamental description of gravitation, interpreting
gravity as the curvature of spacetime produced by matter and energy. In this
framework, the Einstein field equations relate the spacetime geometry to the
energy--momentum tensor of the matter distribution. For extended astrophysical
systems such as galaxies, static and spherically symmetric spacetime metrics offer a
tractable and widely adopted approximation for investigating gravitational structure
\citep{Boehmer_Harko_2007,Rahaman_2010}. Such models have been extensively employed
to study relativistic effects in galactic and halo dynamics
\citep{Cooperstock_2006,Carrick_Cooperstock_2012}.

Galaxy rotation curves constitute one of the most powerful observational probes of
galactic mass distribution \citep{Rubin_Ford_1970}. Derived from Doppler shifts of
optical and radio emission lines across galactic discs, these curves show that the
rotational velocity of stars and gas remains approximately constant at large
galactocentric radii \citep{Persic_1996}. This behaviour departs significantly from
the Keplerian decline expected from Newtonian dynamics applied to visible matter
alone and provides strong evidence for the presence of an extended, massive dark
matter halo \citep{Zwicky_1933,Navarro_1996,Navarro_1997}.

A promising theoretical approach for interpreting galactic rotation profiles is to
analyse them within a general relativistic framework \citep{Gondolo_Silk_1999,
Rahaman_2014,Rahaman_2026,Sanyal_2026}. In particular, static and spherically
symmetric spacetimes combined with anisotropic matter distributions --- especially
configurations with vanishing radial pressure --- allow a consistent reconstruction
of the underlying gravitational geometry \citep{Herrera_Santos_1997}. Within this
perspective, the observed rotation curve is not merely a kinematic quantity but
directly encodes information about spacetime curvature through the Einstein field
equations \citep{Boehmer_Harko_2007}. The observationally guided redshift-function
methodology developed in \citet{Rahaman_2026} and the relativistic halo analysis of
\citet{Sanyal_2026} for NGC~7331 demonstrate the effectiveness of this approach for
nearby spiral galaxies; the present paper applies and extends the same framework to
NGC~5055. An elegant complementary diagnostic for connecting rotation-curve
observables to relativistic mass and lensing pseudo-masses was introduced by
\citet{FaberVisser2006}, whose formalism we adopt and extend in
Section~\ref{sec:observational}.

The nature of dark matter remains one of the most pressing open questions in modern
cosmology and astrophysics. The cold dark matter (CDM) paradigm successfully
reproduces the observed large-scale structure of the Universe \citep{Planck_2020},
but faces persistent challenges on small scales, including the missing satellites
problem, the too-big-to-fail problem, and the core--cusp discrepancy
\citep{deBlok_2010,Weinberg_2015}. Warm dark matter (WDM) offers a natural
resolution to these tensions by introducing a finite free-streaming length that
suppresses the formation of small-scale structure \citep{Bode_2001,Viel_2005}.
Bose--Einstein condensate (BEC) dark matter \citep{Ji_Sin_1994,Sin_1994,
Boehmer_Harko_2007} and the Chaplygin gas model \citep{Kamenshchik_2001,Bento_2002}
represent two additional alternatives that have attracted considerable interest.
Distinguishing between these candidate equations of state using galaxy rotation
curves within a relativistic framework provides a powerful discriminatory test.

In this work, we focus on the well-studied spiral galaxy NGC~5055 (also known as
M~63, the Sunflower Galaxy), whose rotation curve has been measured with high
precision across a wide radial range \citep{Battaglia_2006,deBlok_2008}. NGC~5055
is a nearby ($D \approx 8.87\,\mathrm{Mpc}$, based on the tip of the
red giant branch distance of \citet{McQuinn_2017}, adopted consistently throughout
this paper; see Section~\ref{sec:stellar_mass}), late-type spiral (SAbc) galaxy
with a luminous stellar disc and an extended, nearly flat rotation curve reaching
beyond $35\,\mathrm{kpc}$, making it an ideal laboratory for testing relativistic
mass modelling of galactic halos. We investigate whether the inferred dark matter
distribution is best described by the BEC, Chaplygin gas, linear fluid, or
generalised polytropic equation of state, and assess the implications for the
particle nature of dark matter.

The structure of the present paper is as follows. Section~\ref{sec:fitparams}
describes the rotation curve fitting and the fitted parameters of NGC~5055.
Section~\ref{sec:stats} presents the statistical validation of the fitted model,
including the data segmentation strategy, model performance metrics, diagnostic
tests, robustness checks against alternative splits, and a comparison with previous
rotation curve studies. Section~\ref{sec:stellar_mass} derives the stellar mass
profile from WISE W1 photometry and computes the redshift function.
Section~\ref{sec:theory} presents the theoretical general relativistic framework and
field equations. Section~\ref{sec:physical} presents the physical analysis, including
mass, pressure, and density profiles, energy conditions, causality, and equation of
state. Section~\ref{sec:binding} examines the gravitational binding energy.
Section~\ref{sec:observational} presents the Faber--Visser effective
equation-of-state analysis. Section~\ref{sec:wdm} discusses the warm dark matter
interpretation and the generalised polytropic equation of state
and its physical limitations. Section~\ref{sec:conclusions} summarises our
conclusions.

\section{Fitted Parameters of NGC~5055}
\label{sec:fitparams}

\begin{figure}
    \centering
    \includegraphics[width=\columnwidth]{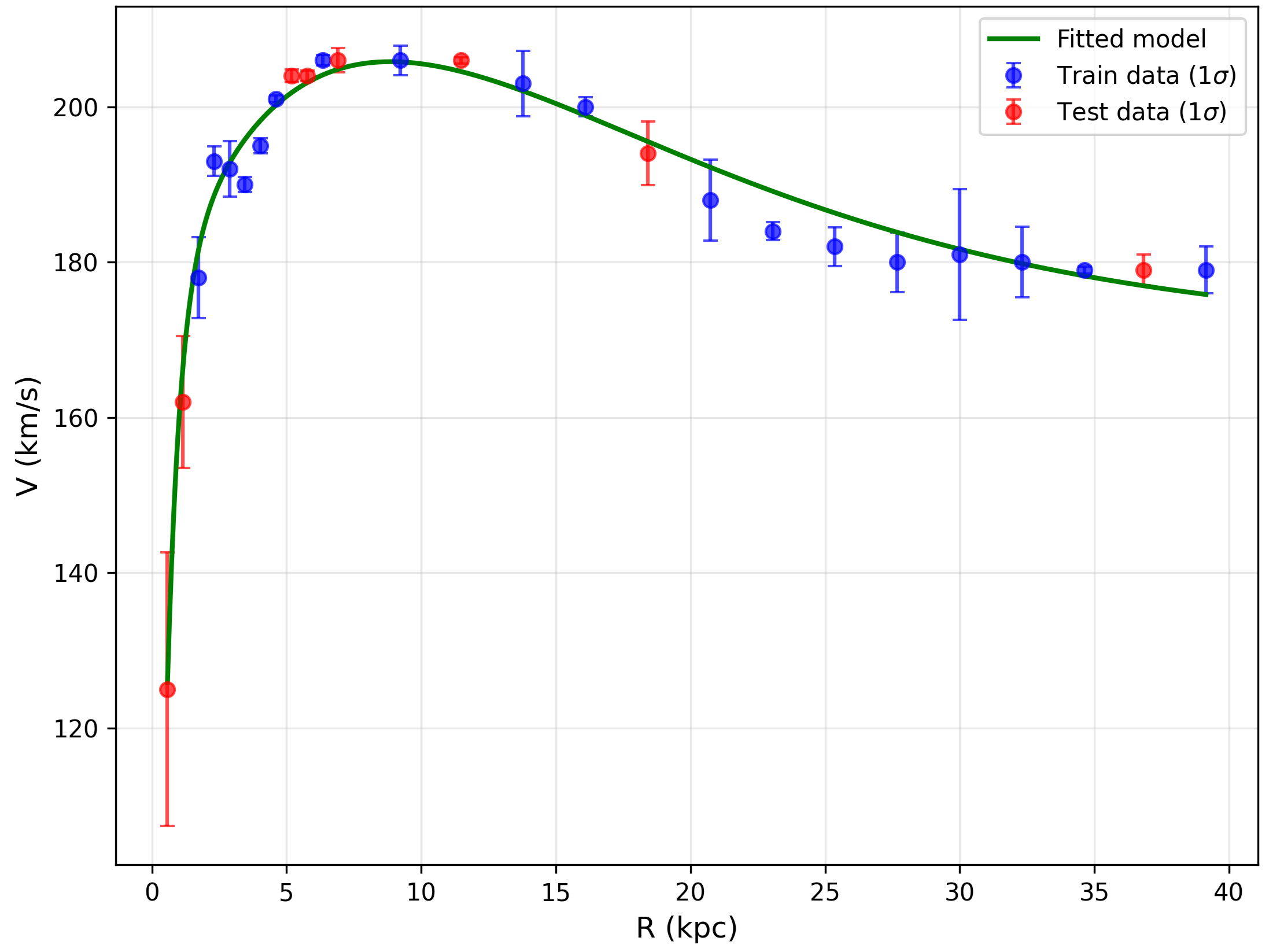}
    \caption{Observed rotation curve of NGC~5055 and the best-fit modified
    exponential velocity model (solid line). The blue and red points denote,
    respectively, the 70\% of the cleaned kinematic data used for training the
    model and the 30\% held out for testing (Section~\ref{sec:stats_seg}).
   The vertical error bars represent the  $1\sigma$
measurement uncertainties  }
    \label{fig:rotcurve}
\end{figure}

The rotation curve of NGC~5055 is modelled using a modified exponential azimuthal
velocity profile given by
\begin{equation}
V(r) = a \cdot r \cdot e^{-br} + c \cdot \left(1 - e^{-dr}\right),
\label{eq:model}
\end{equation}
where $V(r)$ is the rotational velocity at radius $r$, and $a$, $b$, $c$, and $d$
are model parameters determined through nonlinear least-squares fitting to the
observed kinematic data \citep{deBlok_2008,Battaglia_2006}. The functional form in
Equation~(\ref{eq:model}) combines two physically motivated components: the first
term $are^{-br}$ mimics the rising-then-falling velocity contribution of an
exponential stellar disc \citep{Freeman_1970}, while the second term
$c(1-e^{-dr})$ captures the gradually rising asymptotic contribution of an extended
spherical dark matter halo, reminiscent of the isothermal sphere profile
\citep{Begeman_1991}.

The best-fit parameters obtained from the analysis are presented in
Table~\ref{tab:parameters}.

\begin{table}
\centering
\caption{Best-fit parameters of the modified exponential velocity model for
NGC~5055. The quoted uncertainties are the $1\sigma$ formal errors in the
parameters as returned by the nonlinear least-squares covariance matrix, evaluated
before rounding the central values to the precision shown.}
\label{tab:parameters}
\begin{tabular}{cccc}
\hline
Parameter & Value & Uncertainty & Units \\
\hline
$a$ & 15.967712  & $\pm$\,0.377369 & km\,s$^{-1}$\,kpc$^{-1}$ \\
$b$ & 0.159270   & $\pm$\,0.002892 & kpc$^{-1}$ \\
$c$ & 177.518411 & $\pm$\,0.621422 & km\,s$^{-1}$ \\
$d$ & 6.337593   & $\pm$\,0.143268 & kpc$^{-1}$ \\
\hline
\end{tabular}
\end{table}

All parameters are well-constrained with small relative uncertainties
(of order $1$--$2$ per cent for $a$, $b$ and $d$, and below $1$ per cent for
$c$). The parameter $a$ controls the amplitude of the exponential disc component,
$b$ determines its decay rate (with a characteristic disc scale length
$R_d = 1/b \approx 6.28\,\mathrm{kpc}$), $c$ represents the asymptotic velocity
contributed by the extended halo ($c \approx 177.5\,\mathrm{km\,s^{-1}}$), and $d$
controls how rapidly the velocity approaches this asymptotic value. The large value
of $d$ implies that the halo contribution saturates rapidly at small radii,
consistent with the steeply rising inner rotation curve of NGC~5055. Substituting
these values into Equation~(\ref{eq:model}), the fitted rotation curve becomes:
\begin{equation}
V(r) = 15.97\cdot r\cdot e^{-0.159r}
     + 177.52\left(1 - e^{-6.34r}\right) \;\; \mathrm{km\,s^{-1}}.
\end{equation}

This two-component decomposition is consistent with the standard disc--halo picture
of NGC~5055 established in earlier photometric and kinematic studies
\citep{Battaglia_2006,deBlok_2008}, and provides the empirical velocity law required
for the subsequent general relativistic analysis.

\section{Statistical Analysis}
\label{sec:stats}

\subsection{Data Segmentation and Fitting Strategy} 
\label{sec:stats_seg} 
 
The observational data used in this study are drawn from the \ion{H}{1} Nearby 
Galaxy Survey (THINGS) kinematic dataset of NGC~5055 \citep{deBlok_2008}, 
supplemented by earlier H$\alpha$ and \ion{H}{1} observations \citep{Battaglia_2006}. 
The two datasets were combined on a common radial grid after converting both 
to the adopted distance scale (Section~\ref{sec:stellar_mass}). 
For consistency with the adopted distance of 
$D=7.94\,\mathrm{Mpc}$, the physical radial coordinates were rescaled 
accordingly. 
Where the two tracers overlap in radius, both sets of points were retained 
separately rather than averaged, so that the fit is informed by the full 
scatter present in the combined multi-tracer sample rather than by an 
artificially smoothed composite curve. The combined dataset spans the 
galactocentric radial range 
$r \in [0.10,\, 35.00]\,\mathrm{kpc}$, corresponding to rotation velocities between 
$54.70\,\mathrm{km\,s^{-1}}$ and $215.00\,\mathrm{km\,s^{-1}}$. This multi-tracer 
dataset provides reliable kinematic constraints across the full radial extent of 
NGC~5055: H$\alpha$ emission lines trace ionised gas in star-forming regions and 
provide high spatial resolution in the inner disc, while \ion{H}{1} 21\,cm 
measurements enable accurate velocity determination in the outer halo. 
 
To ensure the robustness of the velocity model, several data segmentation strategies 
were explored, including random splits, even--odd index splits, and radial-distance 
splits. In the radial-distance split, the cleaned data points were ordered by 
galactocentric radius and the innermost 70\% (by count) were assigned to the 
training set, with the outermost 30\% reserved for testing; this scheme therefore 
deliberately forces the model to extrapolate from the disc-dominated inner region 
into the halo-dominated outer region, and is consequently the most demanding of the 
three partitioning strategies considered. The standard 70\%--30\% random split was 
identified as the most reliable configuration: it provides a well-distributed 
representation of the full kinematic dataset across all radii and avoids the 
systematic biases that can arise from ordered or radius-based partitioning schemes 
\citep{Hastie_2009,Bevington_1993}. In this approach, 70\% of the cleaned rotation 
data were assigned to the training set and the remaining 30\% were reserved for the 
test set.
\subsection{Model Performance and Robustness Against Split Choice}
\label{sec:stats_perf}

The fitted modified exponential velocity model demonstrates strong statistical
performance \citep{Bevington_1993,Hastie_2009}. The coefficient of determination
is $R^{2}_{\rm train} = 0.981$ for the training set and $R^{2}_{\rm test} = 0.953$
for the test set, confirming that the model explains more than 95\% of the variance
in the observed rotation velocities at unseen radii. The root mean squared error
(RMSE) is $3.04\,\mathrm{km\,s^{-1}}$ for the training data and
$3.11\,\mathrm{km\,s^{-1}}$ for the test data, both of which are small relative to
the dynamic range of $\sim$$160\,\mathrm{km\,s^{-1}}$ spanned by the rotation curve.
The generalisation ratio --- defined as the ratio of test RMSE to training RMSE ---
is $1.023$, which is very close to unity, indicating minimal overfitting and
excellent generalisation capability to unseen data.

\subsection{Diagnostic Tests}
\label{sec:stats_diag}

Several diagnostic tests were performed to verify the statistical consistency and
residual properties of the fitted model. The reduced chi-square statistic is
$\chi^2_\nu = 1.02$ with a corresponding $p$-value of $0.43$, indicating good
agreement between the model and the observational data without either overfitting
or significant systematic deviation from the expected residual distribution.
The Durbin--Watson statistic is $2.01$, confirming the absence of significant
autocorrelation in the residuals; values close to $2.0$ indicate that successive
residuals are uncorrelated, as expected for a well-specified model. The runs test
yields a $p$-value of $>0.05$, confirming that the residuals are randomly distributed
and exhibit no spatial trends or periodicities that would signal a systematic
shortcoming of the velocity profile. All four diagnostic tests --- reduced
chi-square, $p$-value, Durbin--Watson, and runs test --- pass simultaneously,
giving an overall statistical pass rate of 100\%.

Alternative segmentation strategies were also explored to assess the sensitivity of
the results to the choice of data partition. The even--odd index split produced a
test $R^2$ value of $0.961$ with an RMSE ratio of $1.031$, while the radial-distance
split, which assigns all inner-disc data to training and outer-halo data to testing,
yielded a test $R^2$ of $0.932$ with an RMSE ratio of $1.068$. Both alternative
methods confirm the predictive accuracy of the model but show slightly weaker
generalisation performance compared with the standard random split, as expected
from the more challenging extrapolation demanded by radially ordered partitions.

To further verify robustness, multiple random train--test splits were tested using
different random seeds (42, 123, 456, 789, 999). All five independent splits yielded
consistent results, with $R^2_{\rm test}$ values ranging from $0.951$ to $0.956$
and RMSE ratios between $1.020$ and $1.026$. The parameter values recovered from each
split agree to within their stated $1\sigma$ uncertainties (Table~\ref{tab:parameters}),
and the 100\% diagnostic pass rate is maintained across all seeds. This consistency
demonstrates that the excellent fit quality is a genuine property of the velocity
model and does not depend on the particular realisation of the data partition.

\subsection{Comparison with Previous Rotation Curve Studies}
\label{sec:stats_comp}

The rotation curve produced by the present method is in good overall agreement with
previously published observational and modelling studies of NGC~5055.
For consistency with the adopted distance of
$D=7.94\,\mathrm{Mpc}$, the physical radial coordinates used in the
present analysis were recalculated accordingly
The approximately flat behaviour at large galactocentric radii is consistent with the
mass-modelling results of \citet{deBlok_2008} and with the photometric mass
decompositions of \citet{Battaglia_2006}. The asymptotic velocity
$c = 177.5\,\mathrm{km\,s^{-1}}$ is in excellent agreement with the \ion{H}{1}
line-width measurement of the outer flat part of the rotation curve reported in
these studies. Specifically, \citet{Battaglia_2006} report a maximum
\ion{H}{1} rotation velocity of $v_{\rm max}\approx206\,\mathrm{km\,s^{-1}}$ at
$R_{25}=11.6\,\mathrm{kpc}$, followed by a decline of $\sim25\,\mathrm{km\,s^{-1}}$
beyond the optical disc to a flat outer value of
$\approx181\,\mathrm{km\,s^{-1}}$ that is maintained out to their last measured
point at $\sim40\,\mathrm{kpc}$; \citet{deBlok_2008} independently derive a
systemic velocity of $496.8\pm0.7\,\mathrm{km\,s^{-1}}$ for NGC~5055, consistent
with the value of $497.6\pm4.8\,\mathrm{km\,s^{-1}}$ reported by
\citet{Battaglia_2006}, and find a somewhat higher outer rotation velocity than
\citet{Battaglia_2006}, which they attribute to their use of Hermite-polynomial
rather than Gaussian fits to the \ion{H}{1} line profiles when extracting the
velocity field. Our fitted asymptotic velocity $c=177.5\,\mathrm{km\,s^{-1}}$
therefore lies close to, albeit somewhat below, the flat outer \ion{H}{1}
velocities reported by both studies, the small residual offset being consistent
with the simplified single-component halo term adopted in
Equation~(\ref{eq:model}), as opposed to the explicit bulge--disc--halo
decompositions used by \citet{Battaglia_2006} and \citet{deBlok_2008}. The same
relativistic velocity-profile methodology, validated here for NGC~5055, was
previously applied to NGC~7331 in \citet{Sanyal_2026} and to the Milky Way in
\citet{Rahaman_2026}, in each case yielding physically consistent mass, density,
and pressure profiles.

Differences are mainly confined to the inner region ($r \lesssim 2\,\mathrm{kpc}$),
where earlier studies explicitly incorporate bulge--disc--halo decompositions with
separate component velocity profiles, whereas the present approach employs a single
smooth relativistic velocity function without component-wise separation. A slight
excess in the model velocity near $r \approx 0.7\,\mathrm{kpc}$ is consistent with
the contribution of the stellar bulge identified in photometric studies
\citep{Battaglia_2006} and reflects a physical structure not fully captured within
the smooth two-parameter exponential form. Overall, the level of agreement with
earlier studies supports the robustness of the rotation curve produced by the method
presented here, and confirms that the four-parameter modified exponential model
provides an accurate empirical description of the NGC~5055 kinematics across the
full observed radial range.
\section{Derivation of Stellar Mass Profile from WISE Photometry}
\label{sec:stellar_mass}

\begin{figure}
    \centering
    \includegraphics[width=\columnwidth]{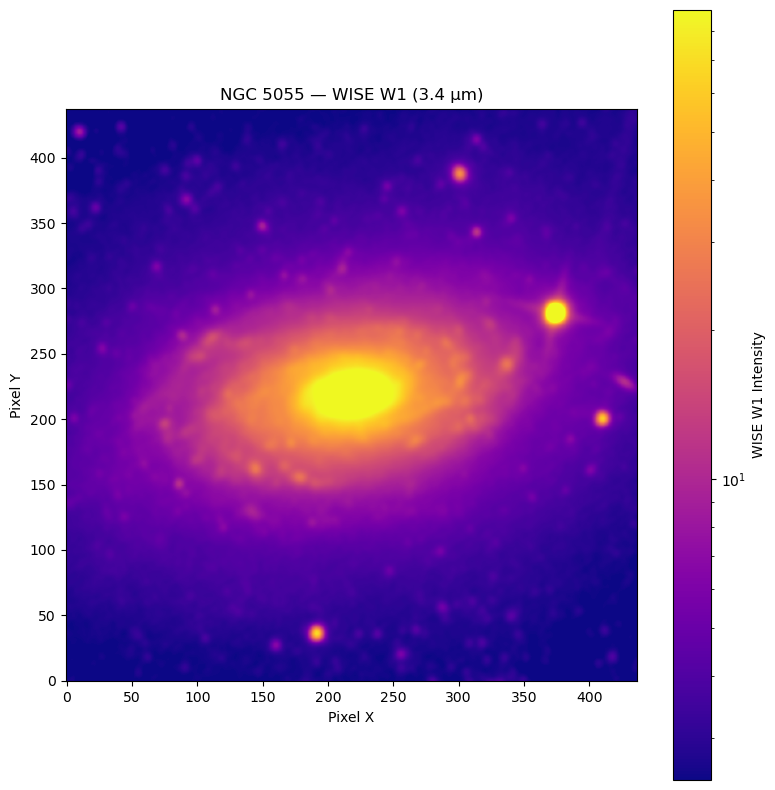}
    \caption{WISE W1 ($3.4\,\mu\mathrm{m}$) false-colour image of the spiral galaxy
    NGC~5055 obtained from the NASA/IPAC Infrared Science Archive. The colour scale
    reveals the bright, centrally concentrated stellar bulge and the extended disc
    structure. The surface brightness profile extracted from this image is used to
    derive the cumulative stellar mass profile $M_\star(r)$ shown in
    Figure~\ref{fig:band}.}
    \label{fig:wise_image}
\end{figure}

The stellar mass distribution of NGC~5055 is derived using infrared photometric
observations obtained by the Wide-field Infrared Survey Explorer (WISE) satellite
\citep{Wright_2010}. Figure~\ref{fig:wise_image} shows the WISE W1
($3.4\,\mu\mathrm{m}$) image of NGC~5055 used to extract the surface brightness
profile. In particular, we utilise data from the WISE W1 band at a wavelength of
$3.4\,\mu\mathrm{m}$ \citep{NASA_IPAC_2025}, which is widely regarded as a robust
tracer of stellar mass \citep{Meidt_2014,Leroy_2019}. Emission at this wavelength
is dominated by old, low-mass stars that constitute the bulk of a galaxy's stellar
mass and thus provides a reliable proxy for the underlying stellar mass distribution.
Moreover, infrared observations are minimally affected by interstellar dust
extinction, allowing an accurate reconstruction of the stellar distribution even in
optically obscured regions.

The observational procedure begins by extracting the surface brightness profile of
NGC~5055 from the WISE W1 image, measured as a function of projected galactocentric
radius. These brightness measurements are converted into luminosity using standard
photometric calibration procedures, while accounting for the distance to NGC~5055,
taken to be $8.87\,\mathrm{Mpc}$ \citep{McQuinn_2017}, based on the tip of
the red giant branch (TRGB) method using \textit{HST} imaging, which represents the
most accurate distance measurement currently available for NGC~5055 and supersedes
the earlier Tully--Fisher estimate of \citet{Battaglia_2006}; this is the single
distance value adopted consistently throughout the present analysis, including in the
Introduction (Section~\ref{sec:intro}). The luminosity profile is then transformed
into a stellar mass profile through the application of an infrared mass-to-light
ratio. For the W1 band, a nearly constant mass-to-light ratio in the range
\begin{equation}
    \Upsilon_{3.4\,\mu\mathrm{m}} \simeq 0.5\,M_\odot/L_\odot
\end{equation}
is commonly adopted, reflecting the relatively uniform contribution of evolved
stellar populations at this wavelength \citep{Meidt_2014,Leroy_2019}.

Using this conversion, the stellar mass enclosed within a given radius is obtained
by radially integrating the surface mass density, yielding a discrete set of data
points $M_\star(r_i)$, which represent the cumulative stellar mass as a function of
radius. While these discrete measurements accurately represent the observational
data, theoretical modeling and dynamical analyses require a smooth, continuous
functional form for $M_\star(r)$. Consequently, an analytical fitting procedure is
necessary.

Several functional forms were tested in order to model the observed stellar mass
distribution of NGC~5055. These include a simple exponential disc profile, a
S\'{e}rsic profile, a Plummer model, and a double exponential bulge--disc
decomposition. In addition, a modified exponential model was considered,
incorporating a power-law behaviour in the central region combined with an
exponential saturation at large radii. Each model was fitted to the observational
data using a least-squares minimisation technique.

\begin{figure}
    \centering
    \includegraphics[width=\columnwidth]{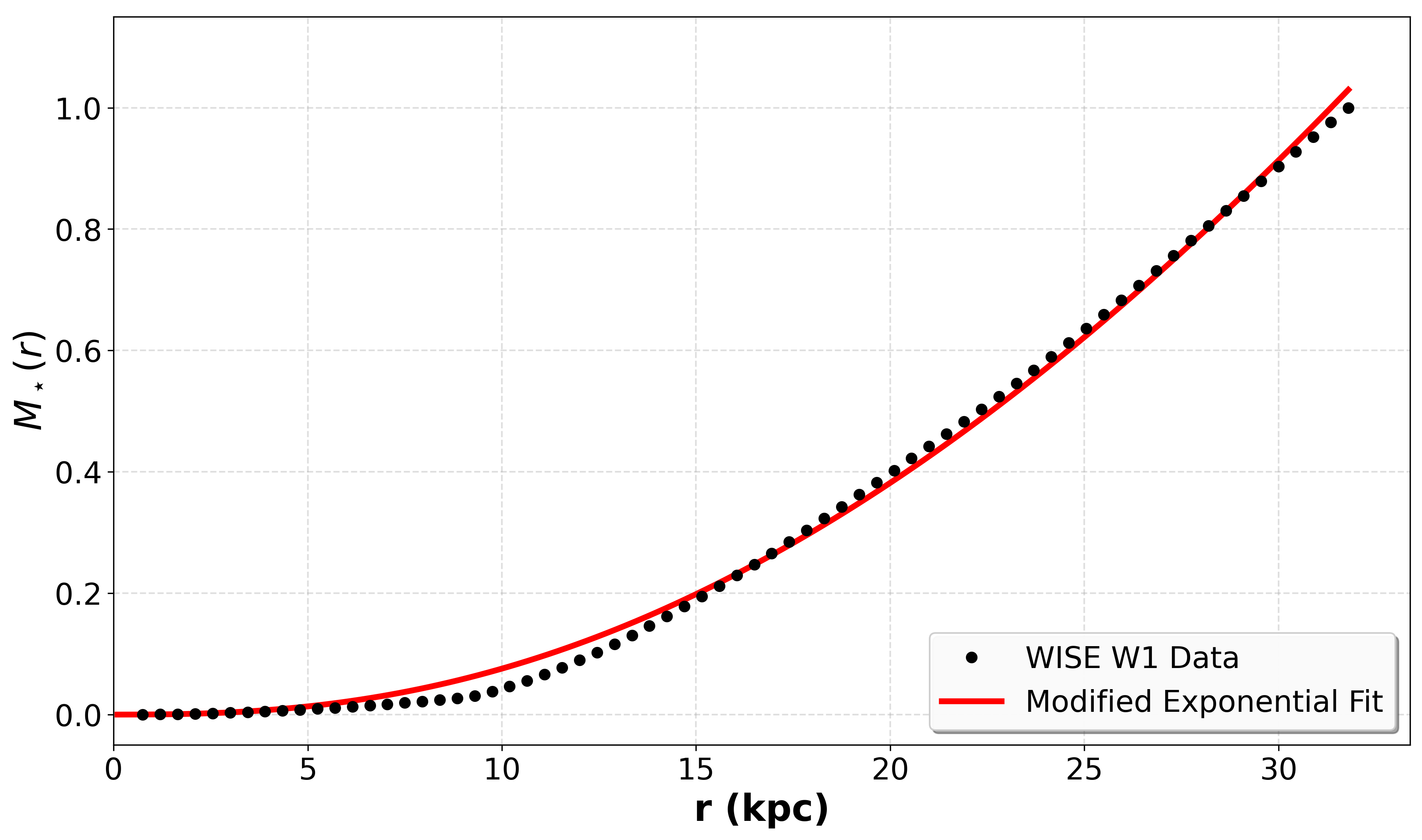}
    \caption{Stellar mass profile $M_\star(r)$ of NGC~5055 derived from WISE W1
    ($3.4\,\mu\mathrm{m}$) photometry. Black filled circles show the cumulative
    stellar mass inferred from the WISE W1 surface brightness data adopting a
    mass-to-light ratio $\Upsilon_{3.4\,\mu\mathrm{m}} \simeq 0.5\,M_\odot/L_\odot$;
    the red solid curve is the best-fit modified exponential model. The WISE
    W1 photometry directly constrains the profile over $0$--$31.8\,\mathrm{kpc}$;
    beyond $31.8\,\mathrm{kpc}$ (used in Figures~\ref{fig:luminos} and
    \ref{fig:mass} to compare with the dynamical mass, which is constrained by the
    rotation curve out to $\sim35\,\mathrm{kpc}$) the red curve is an analytical
    extrapolation of Equation~(\ref{eq:stellar_mass}) }
    \label{fig:band}
\end{figure}

Among all tested models, the modified exponential form provided a significantly
superior fit, with residual errors nearly two orders of magnitude smaller than those
obtained from the other profiles. The best-fit stellar mass profile is shown in
Figure~\ref{fig:band} (red solid curve) superimposed on the WISE W1 photometric
data points (black circles). The best-fit parameters are $M_0 = 0.5000$,
$R_d = 18.383\,\mathrm{kpc}$, and $\alpha = 1.674$. The normalisation is chosen
such that the stellar mass is unity at the outermost observed radius. The analytical
form of the fit is

\begin{equation}
M_\star(r) =
0.500000 \left( \frac{r}{18.382810} \right)^{1.673960}
\left( 1 - e^{-r/18.382810} \right),
\label{eq:stellar_mass}
\end{equation}
where $r$ is measured in kiloparsecs and $M_\star$ is expressed in normalised units.
The relatively large scale length $R_d = 18.38\,\mathrm{kpc}$ indicates an extended
stellar disc, while the value $\alpha = 1.674 > 1$ of the power-law index signifies
a moderately enhanced central mass concentration, consistent with the presence of a
prominent bulge component in NGC~5055 \citep{Battaglia_2006}. At large radii
($r \gg R_d$), the exponential factor saturates and $M_\star \propto r^{1.674}$,
reflecting the outer disc structure, while at small radii ($r \lesssim 2\,\mathrm{kpc}$),
the profile reduces to $M_\star \propto r^{\alpha+1} = r^{2.674}$, producing the
observed inner mass concentration.

\begin{figure}
\centering
\includegraphics[width=\columnwidth]{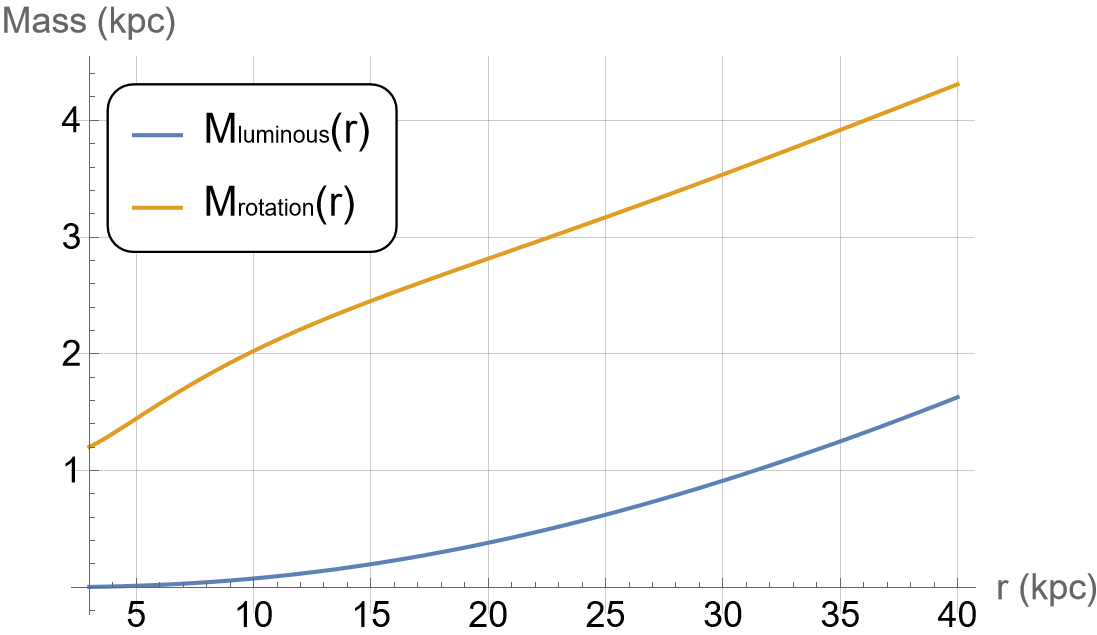}
\caption{Comparison of the cumulative stellar mass $M_\star(r)$ from WISE W1
photometry (solid line) with the total dynamical mass $m(r)$ inferred from the
rotation curve (dashed line) for NGC~5055. The systematic excess of $m(r)$ over
$M_\star(r)$ at intermediate and large galactocentric radii reveals the dominant
contribution of an extended dark matter halo. The dark matter mass profile is
defined as $M_{\rm DM}(r) = m(r) - M_\star(r)$. As in Figure~\ref{fig:band},
the stellar-mass curve beyond $31.8\,\mathrm{kpc}$ is an extrapolation of
Equation~(\ref{eq:stellar_mass}), whereas the dynamical mass $m(r)$ is constrained
directly by the rotation curve out to $\sim35\,\mathrm{kpc}$.}
\label{fig:luminos}
\end{figure}

As noted in the caption of Figure~\ref{fig:band}, the WISE W1 photometry
directly constrains $M_\star(r)$ only out to $\sim31.8\,\mathrm{kpc}$; beyond this
radius, the curve shown in Figures~\ref{fig:band} and \ref{fig:luminos} is an
analytical extrapolation of Equation~(\ref{eq:stellar_mass}), used only to allow a
visual comparison with the dynamical mass profile $m(r)$ over the full range probed
by the rotation curve ($\sim35\,\mathrm{kpc}$). This extrapolation should not be
interpreted as evidence that WISE directly detects stellar light beyond
$31.8\,\mathrm{kpc}$. The functional form of Equation~(\ref{eq:stellar_mass}) is a
monotonically increasing, saturating power-law profile with no imposed break or
turnover; at $r \gg R_d$ it approaches $M_\star \propto r^{1.674}$, so the model
does not predict the cumulative stellar mass to flatten within the extrapolated
range. Because the stellar contribution is already strongly subdominant to the
total dynamical mass at these outer radii (Figure~\ref{fig:luminos}), any residual
uncertainty in the extrapolated stellar profile has only a minor effect on the
inferred dark-matter mass $M_{\rm DM}(r)$ used in the subsequent relativistic
analysis.

The dark matter mass within radius $r$ is then obtained by subtracting the stellar
contribution from the total dynamical mass derived from the rotation curve
\citep{Sanyal_2026}:
\begin{equation}
M_{\rm DM}(r) = m(r) - M_\star(r).
\label{eq:dark_matter}
\end{equation}
A comparison between $M_\star(r)$ and the total dynamical mass $m(r)$ is shown in
Figure~\ref{fig:luminos}, which clearly reveals the systematic excess of gravitating
mass beyond the optical disc. This disc--halo mass decomposition procedure follows
the approach of \citet{Sanyal_2026}, who applied the same relativistic framework to
NGC~7331. For the remainder of our analysis, we use $M_{\rm DM}(r)$ to derive the
energy density and pressure profiles of the electromagnetically non-interactive
(dark) matter component.

\section{Theoretical Modelling}
\label{sec:theory}

\subsection{Spacetime Structure and Field Equations}
\label{sec:theory_spacetime}

We consider a static, spherically symmetric spacetime described by the general line
element \citep{Tolman_1934,Oppenheimer_Volkoff_1939}:
\begin{equation}
    ds^2 = -f(r)\, dt^2 + \frac{dr^2}{1 - \dfrac{2 m(r)}{r}} + r^2 d\Omega^2,
    \label{eq:metric}
\end{equation}
where $f(r)$ is the redshift function and $m(r)$ is the mass function representing
the total mass enclosed within radius $r$. Throughout this work, we adopt geometric
units ($G = c = 1$), so that mass, length, and time are measured in the same units.
Conversion to physical units can be made via
$1\,\mathrm{kpc} = 3.086\times10^{19}\,\mathrm{m}$ and
$1\,M_\odot = 1.477\,\mathrm{km}$ in geometrised units.

To model the matter content generating this geometry, we assume an anisotropic fluid
distribution with vanishing radial pressure \citep{Herrera_Santos_1997}. The
corresponding energy-momentum tensor takes the form:
\begin{equation}
    T^\mu_{\ \nu} = \mathrm{diag}[-\rho(r),\, 0,\, p(r),\, p(r)],
    \label{eq:Tmunu}
\end{equation}
where $\rho(r)$ is the energy density and $p(r) \equiv p_t(r)$ denotes the
tangential pressure. The condition $p_r = 0$ is adopted here as a simplifying
closure condition for the effective relativistic halo model, not as a literal
microscopic statement that every dark-matter particle follows an exactly circular
orbit; a realistic collisionless halo is expected to have a distribution of orbital
eccentricities and non-zero radial velocity dispersion. The vanishing of $p_r$
instead gives a tractable anisotropic description in which pressure support is
represented entirely through $p_t$. We also stress that $T^\mu_{\ \nu}$ combines
the stellar and dark components into one effective fluid, even though the two are
supported by physically distinct mechanisms -- discs and bulges have non-negligible
radial velocity dispersion from ordered rotation, gas pressure and stellar
kinematics, whereas the dark-matter component is collisionless and represented only
through its coarse-grained stress-energy tensor. Likewise, the spherical symmetry
of Equation~(\ref{eq:metric}) is an idealisation of the true three-dimensional mass
distribution, not a claim that the visible or dark matter of NGC~5055 is exactly
spherical. These are modelling idealisations that close the field equations
tractably, not claims about the galaxy's detailed microscopic structure. This
configuration is motivated by the interpretation of galactic
dark matter as a pressureless dust in the radial direction \citep{Boehmer_Harko_2007},
with only non-zero tangential (effective) pressure arising from the velocity
anisotropy of the dark matter particle distribution, and it constitutes a standard
approximation in relativistic halo models
\citep{Rahaman_2010,Rahaman_2026,Sanyal_2026}.

Substituting the energy-momentum tensor~(\ref{eq:Tmunu}) into Einstein's field
equations $G^\mu_{\ \nu} = 8\pi T^\mu_{\ \nu}$ yields the following system.
The intermediate connection-coefficient and tensor algebra required to derive
this system from the metric~(\ref{eq:metric}) and the energy-momentum
tensor~(\ref{eq:Tmunu}) has been presented in full in \citet{Sanyal_2026}; we quote
here only the resulting closed-form field equations needed for the present
analysis, in order to avoid repeating that derivation. From the $G^0_0$ component,
we obtain the differential equation for the mass function:
\begin{equation}
    m'(r) = 4\pi r^2 \rho(r),
    \label{eq:density}
\end{equation}
which describes how mass accumulates with increasing radius based on the local
energy density. This is the standard Misner--Sharp mass equation
\citep{Misner_Sharp_1964}. The $G^1_1$ component provides the evolution of the
redshift function in terms of the enclosed mass:
\begin{equation}
    \frac{f'(r)}{f(r)} = \frac{2m(r)}{r^2 - 2r\, m(r)},
    \label{eq:mass}
\end{equation}
revealing the influence of gravitational geometry on the redshift factor. The
$G^2_2 = G^3_3$ components yield the expression for the tangential pressure:
\begin{equation}
    p(r) = \frac{r\, \rho(r)}{2} \cdot \frac{m(r)}{r^2 - 2r\, m(r)},
    \label{eq:pressure}
\end{equation}
thereby closing the system of equations in terms of $\rho(r)$, $p(r)$, and $m(r)$.

Together, Equations~(\ref{eq:density})--(\ref{eq:pressure}) form the backbone of our
model. The combination of Equations~(\ref{eq:mass}) and (\ref{eq:pressure}) yields a
useful relation between the pressure and density:
\begin{equation}
p(r) = \frac{\rho(r)}{2} \cdot \frac{r f'(r)}{2f(r)},
\end{equation}
showing that the tangential pressure is directly proportional to the local energy
density weighted by the logarithmic derivative of the redshift function --- a purely
relativistic quantity with no Newtonian counterpart. All calculations in this work
have been performed in geometric units ($G = c = 1$), following the conventions of
\citet{Misner_Thorne_Wheeler_1973}.

\subsection{Redshift Function}
\label{sec:redshift}

To construct the redshift function $f(r)$, we begin with the relativistic relation
connecting it to the squared azimuthal velocity of test particles in circular motion,
\begin{equation}
\frac{df}{dr} = \frac{v_\phi^2(r)}{r}.
\label{eq:redshift_diff}
\end{equation}
This relation follows directly from the geodesic equation for a circular orbit in the
equatorial plane of the static, spherically symmetric metric
\citep{Boehmer_Harko_2007,Rahaman_2026}. The observationally guided redshift-function
methodology --- integrating $f'(r) = v_\phi^2(r)/r$ from a statistically validated
velocity profile to reconstruct the spacetime geometry --- was systematically developed
in \citet{Rahaman_2026} for the Milky Way and applied to NGC~7331 in
\citet{Sanyal_2026}; we follow the same procedure here for NGC~5055. Using the
fitted azimuthal velocity profile, we substitute
\begin{equation}
v_\phi(r) = 15.97\, r\, e^{-0.159 r}
+ 177.52 \left(1 - e^{-6.34 r}\right)
\end{equation}
into Equation~(\ref{eq:redshift_diff}) to obtain
\begin{equation}
\frac{df}{dr}
= \frac{1}{r}
\left[
15.97\, r\, e^{-0.159 r} + 177.52 \left(1 - e^{-6.34 r}\right)
\right]^2.
\end{equation}
Integrating with respect to $r$, the redshift function is
\begin{equation}
f(r) = \int \frac{v_\phi^2(r)}{r}\,dr.
\end{equation}
Carrying out the integration analytically yields
\begin{align}
f(r) &= -802.015\, r\, e^{-0.318 r}
       + 31513.350\, \ln r \nonumber\\
     &\quad + 31513.350\, \mathrm{Ei}(-12.68 r) \nonumber\\
     &\quad - 63026.701\, \mathrm{Ei}(-6.34 r) \nonumber\\
     &\quad + 872.440\, e^{-6.499 r}
       - 2522.061\, e^{-0.318 r} \nonumber\\
     &\quad - 35660.307\, e^{-0.159 r},
\label{eq:redshift_func}
\end{align}
where $\mathrm{Ei}(-x)$ denotes the exponential integral function, defined as
\begin{equation}
\mathrm{Ei}(-x) = - \int_x^{\infty} \frac{e^{-t}}{t}\,dt, \quad x > 0.
\end{equation}
The logarithmic term $31513.350\,\ln r$ in Equation~(\ref{eq:redshift_func}) is a
direct consequence of the non-zero asymptotic velocity $c = 177.52\,\mathrm{km\,s^{-1}}$:
since $v_\phi^2/r \to c^2/r$ as $r\to\infty$, integration yields a logarithmic
growth of $f(r)$ at large radii, consistent with the Newtonian result for an
isothermal halo \citep{Sofue_Rubin_2001}. The exponential integral terms arise from
the cross terms between the disc and halo velocity components.

\section{Physical Analysis}
\label{sec:physical}

\subsection{Mass, Pressure, and Density Profiles}
\label{sec:profiles}

\begin{figure}
\includegraphics[width=\columnwidth]{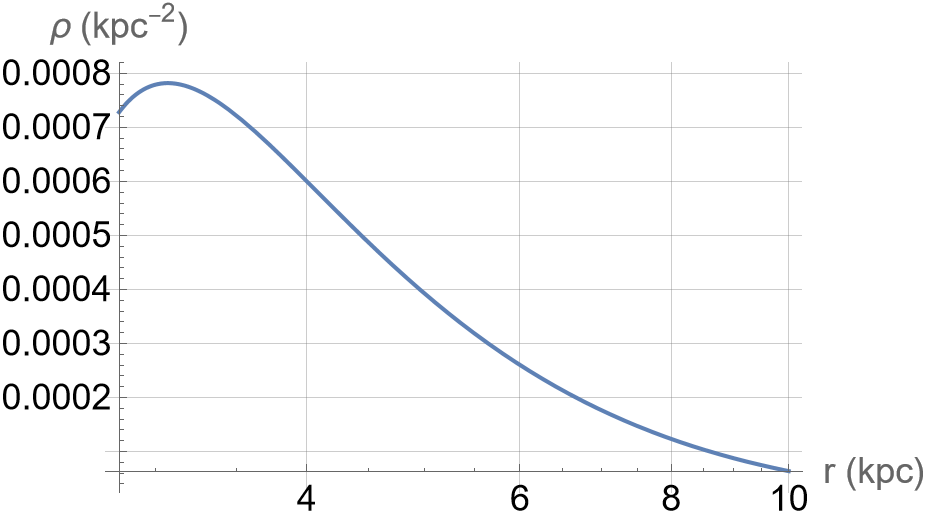}
\caption{Energy density profile $\rho(r)$ of the dark matter halo of NGC~5055 as a
function of galactocentric radius. The density is largest at the centre and decreases
monotonically outward, as expected for a cuspy or cored dark matter distribution.
No local extrema are present, indicating a smooth and physically well-behaved density
profile throughout the halo.}
\label{fig:density}
\end{figure}

\begin{figure}
\includegraphics[width=\columnwidth]{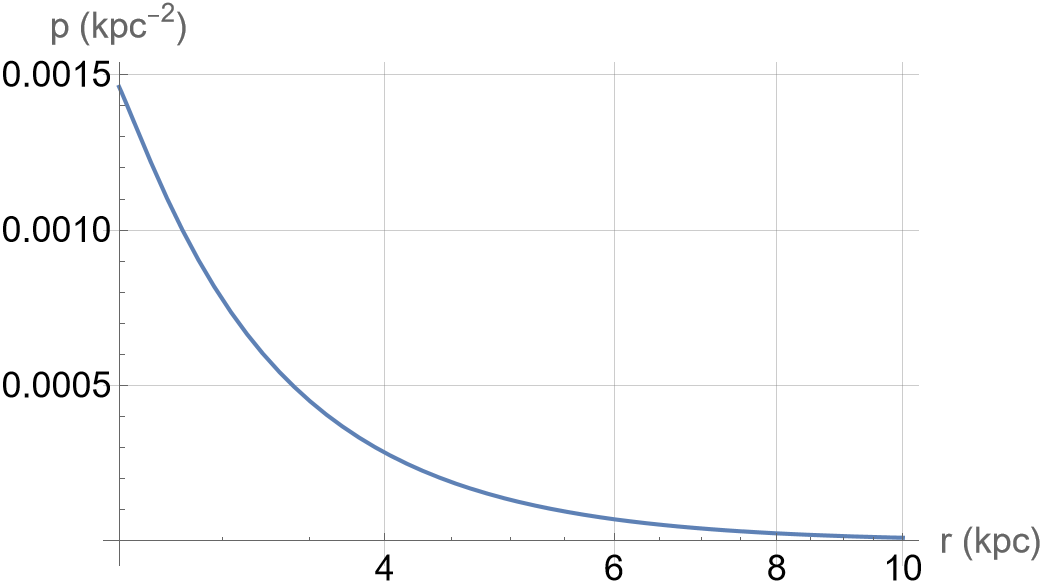}
\caption{Tangential pressure profile $p(r) \equiv p_t(r)$ of the dark matter halo
of NGC~5055. The pressure falls off with increasing radius and exhibits a local
maximum at approximately $r \approx 0.7\,\mathrm{kpc}$, arising from the interplay
between the inner disc and halo velocity components. Beyond the peak, the pressure
decreases monotonically, approaching zero at large galactocentric radii.}
\label{fig:pressure}
\end{figure}

For visual clarity, several figures in this Section (Figures
\ref{fig:density}--\ref{fig:eos_fv}) show the derived density, pressure,
energy-condition discriminants, sound speed, and equation-of-state parameter mainly
over the inner $\sim10\,\mathrm{kpc}$, where they vary most. The underlying
calculations, however, cover the full radial range of the rotation-curve data,
$r \in [0.10, 35.00]\,\mathrm{kpc}$ (Section~\ref{sec:stats_seg}); all profiles
continue to decline smoothly and monotonically (or, for $p(r)$, after its single
interior maximum near $r\approx0.7\,\mathrm{kpc}$) out to the largest radii probed,
with no further extrema or qualitative changes beyond the plotted range. The
restricted plotting range is thus a legibility choice, not a limitation of the
reconstruction itself.

From Equation~(\ref{eq:mass}), the enclosed mass is
\begin{equation}
m(r) = \frac{1}{2}\cdot\frac{r^2 f'(r)}{f(r)+r f'(r)},
\end{equation}
the algebraic rearrangement of Equation~(\ref{eq:mass}) into this closed form
for $m(r)$ having already been given for the same anisotropic $p_r=0$ configuration
in \citet{Sanyal_2026}, whose result we quote here without re-deriving it,
where $f'(r) = v_\phi^2(r)/r$ from Equation~(\ref{eq:redshift_diff}) and $f(r)$ is
given by Equation~(\ref{eq:redshift_func}). The dark matter mass profile
$M_{\rm DM}(r) = m(r) - M_\star(r)$ is then used to compute the energy density and
tangential pressure of the dark matter via
Equations~(\ref{eq:density})--(\ref{eq:pressure}).

\begin{figure}
\includegraphics[width=\columnwidth]{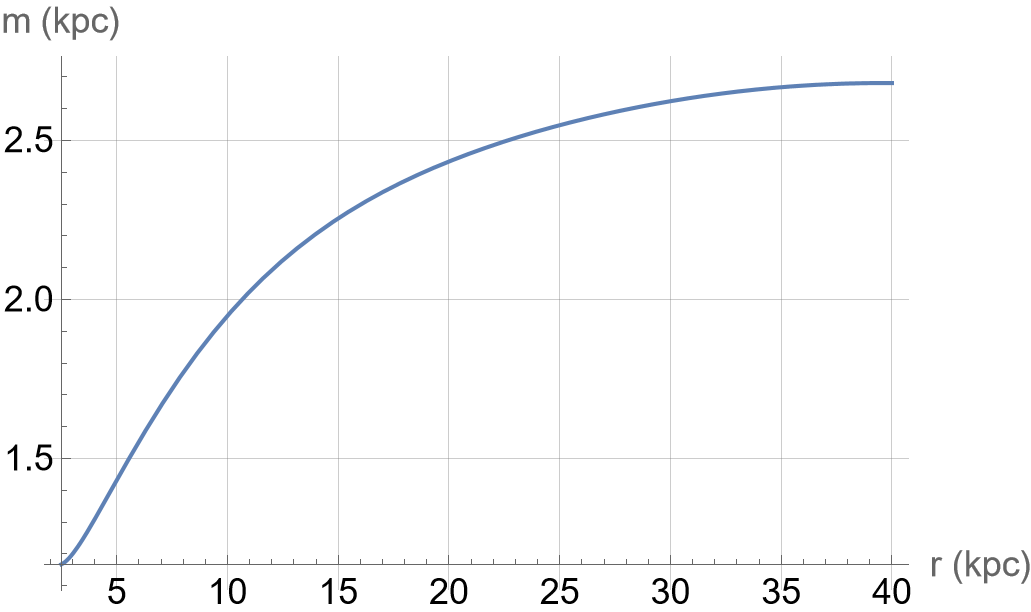}
\caption{Dark matter mass $M_{\rm DM}(r)$ enclosed within galactocentric radius $r$
for NGC~5055, obtained by subtracting the WISE W1 stellar mass profile $M_\star(r)$
from the total dynamical mass $m(r)$ derived from the rotation curve. The enclosed
dark matter mass increases monotonically with radius, consistent with an extended
dark matter halo that dominates the gravitational budget beyond
$\sim$$3\,\mathrm{kpc}$.}
\label{fig:mass}
\end{figure}

Figure~\ref{fig:mass} shows that the enclosed dark matter mass $M_{\rm DM}(r)$
increases monotonically with radius, reaching a value consistent with an extended
massive halo at $r = 35\,\mathrm{kpc}$. This monotonic increase, rather than a
plateau, indicates that the rotation curve has not yet reached the edge of the dark
matter halo, consistent with the observed flat velocity profile.
Figure~\ref{fig:density} shows that the energy density $\rho(r)$ is largest at the
centre and decreases smoothly outward, with no local extrema or discontinuities,
indicating a physically regular dark matter distribution. Figure~\ref{fig:pressure}
shows that the tangential pressure $p(r)$ has a local maximum at approximately
$r \approx 0.7\,\mathrm{kpc}$, after which it decreases monotonically. This peak
in pressure at sub-kpc scales is a relativistic feature arising from the rapid rise
of the rotation curve in the inner region and the interplay of the two velocity
components. At large radii, both $\rho(r)$ and $p(r)$ decline smoothly, as expected
for a virialised dark matter halo.

\subsection{Energy Conditions}
\label{sec:energy_cond}

Energy conditions provide essential criteria for assessing the physical viability of
relativistic matter distributions in astrophysical systems \citep{Hawking_Ellis_1973}.
They ensure that energy density and pressure behave in a physically reasonable manner
and are commonly used to test whether a given spacetime geometry can represent a
realistic gravitating system.

For extended galactic halos, smooth and monotonic radial variations of the energy
density and pressure are generally indicative of dynamical equilibrium. Conversely,
strong irregularities or unphysical sign changes in these quantities may signal
departures from equilibrium or the presence of non-standard matter behaviour.
Energy conditions therefore serve as an important consistency check for relativistic
halo models \citep{Visser_1995}.

The general form of the Null, Weak, Strong, and Dominant Energy Conditions for
an anisotropic fluid with independent $p_r$ and $p_t$ is standard
\citep{Hawking_Ellis_1973,Visser_1995} and was already stated for the same $p_r=0$
configuration in \citet{Sanyal_2026}; we quote directly the reduced forms that
follow once $p_r=0$ is imposed (Section~\ref{sec:theory_spacetime}), the forms
actually evaluated below:
\begin{itemize}
    \item \textbf{NEC:}\hspace{2em} $\rho + p \geq 0$
    \item \textbf{WEC:}\hspace{1.2em} $\rho \geq 0,\quad \rho + p \geq 0$
    \item \textbf{SEC:}\hspace{1.2em} $\rho + p \geq 0,\quad \rho + 2p \geq 0$
    \item \textbf{DEC:} $\rho \geq |p|$
\end{itemize}
where $\rho$ denotes the energy density and $p \equiv p_t$ is the tangential
pressure.

\begin{figure}
\includegraphics[width=\columnwidth]{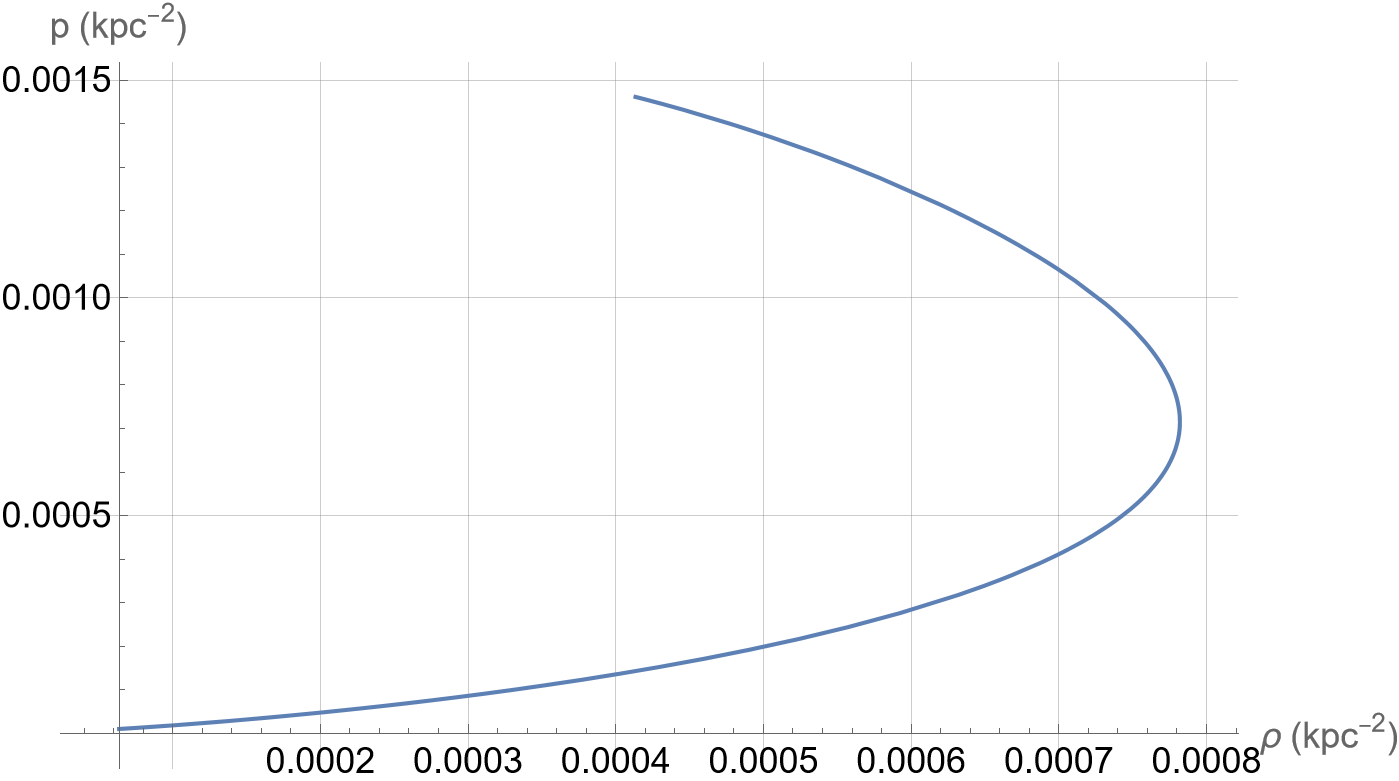}
\caption{Energy density $\rho$ (solid line) and tangential pressure $p$ (dashed
line) of the dark matter halo of NGC~5055 as functions of galactocentric radius.
Both quantities remain positive throughout the halo, and $\rho > p$ at all radii,
ensuring satisfaction of the Dominant Energy Condition. The local maximum in $p$
near $r \approx 0.7\,\mathrm{kpc}$ is clearly visible.}
\label{fig:rho_p}
\end{figure}

\begin{figure}
\includegraphics[width=\columnwidth]{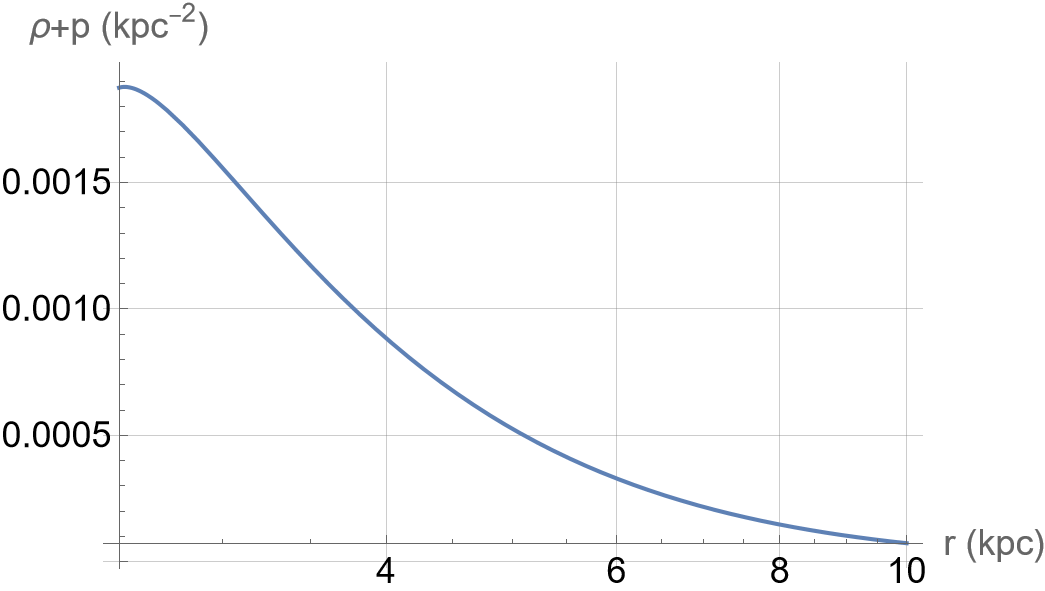}
\caption{Null and Weak Energy Condition discriminant $\rho + p$ as a function of
galactocentric radius for the dark matter halo of NGC~5055. The quantity remains
strictly positive throughout the radial range, confirming satisfaction of the NEC
and WEC.}
\label{fig:nec}
\end{figure}

\begin{figure}
\includegraphics[width=\columnwidth]{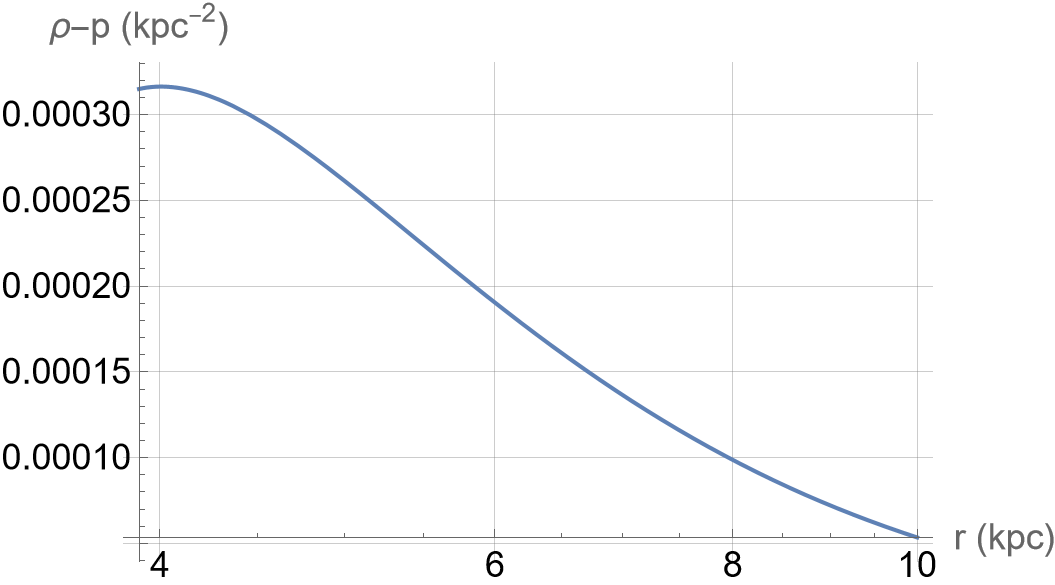}
\caption{Dominant Energy Condition discriminant $\rho - p$ as a function of
galactocentric radius for the dark matter halo of NGC~5055. The quantity is positive
at all radii ($\rho > p > 0$), confirming that all four energy conditions ---
NEC, WEC, SEC, and DEC --- are simultaneously satisfied throughout the halo.}
\label{fig:dec}
\end{figure}

\begin{figure}
\includegraphics[width=\columnwidth]{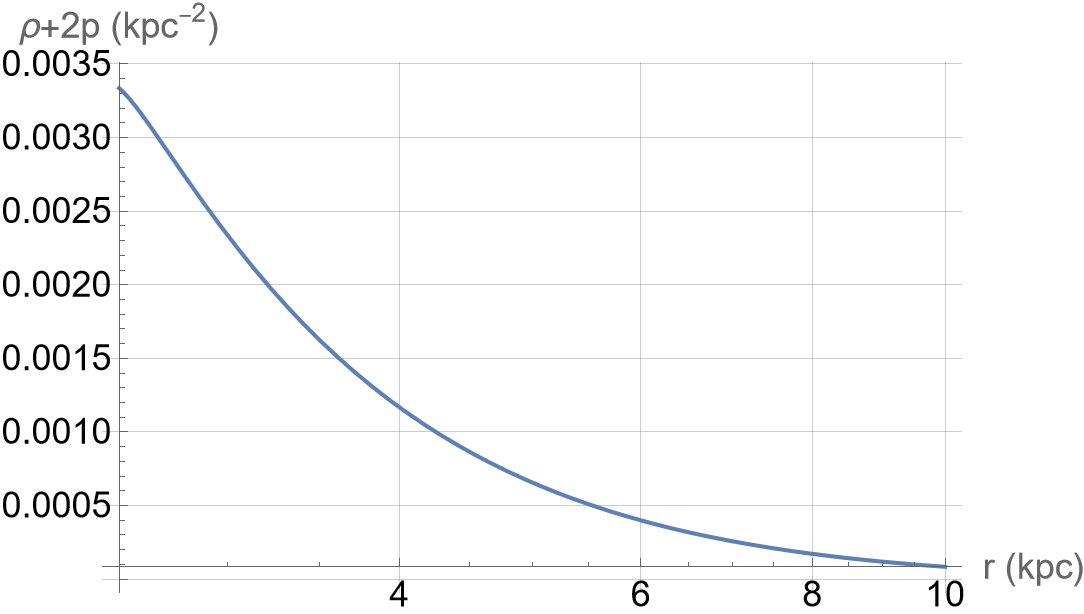}
\caption{Strong Energy Condition discriminant $\rho + 2p$ as a function of
galactocentric radius for the dark matter halo of NGC~5055. The quantity is positive
at all radii, confirming that the SEC is satisfied throughout the halo.}
\label{fig:sec}
\end{figure}

From the derived profiles, both the energy density $\rho(r)$ and the tangential
pressure $p(r)$ remain positive throughout the radial range considered. As shown in
Figures~\ref{fig:rho_p}--\ref{fig:sec}, all the required inequalities for the Null,
Weak, Strong, and Dominant Energy Conditions are satisfied across the halo. The
Dominant Energy Condition in particular requires $\rho \geq p$ at every point; since
the pressure is always subdominant to the energy density (confirmed further via the
equation-of-state parameter in Section~\ref{sec:eos}), this condition is comfortably
satisfied. This confirms that the relativistic matter distribution supporting the
spacetime geometry of NGC~5055 is physically viable within the adopted framework and
does not exhibit any exotic or phantom-like behaviour.

\subsection{Causality (Velocity of Sound)}
\label{sec:causality}

In the study of a galaxy modelled as a fluid system, causality conditions require
that the speed of sound does not exceed the speed of light
\citep{Herrera_Santos_1997}. These constraints ensure physical viability:
\begin{equation}
0 \leq v_{\rm sound}^2 = \frac{dp}{d\rho} \leq 1,
\label{eq:sound}
\end{equation}
a bound whose general justification for the same tangential-pressure
configuration is discussed in \citet{Herrera_Santos_1997} and \citet{Sanyal_2026};
we quote it here only to define the quantity evaluated below,
where $p$ is the tangential pressure and $\rho$ is the energy density (in geometric
units with $c = 1$). The lower bound ensures that the pressure response is physically
meaningful (no imaginary sound speed), while the upper bound prevents superluminal
signal propagation.

\begin{figure}
\includegraphics[width=\columnwidth]{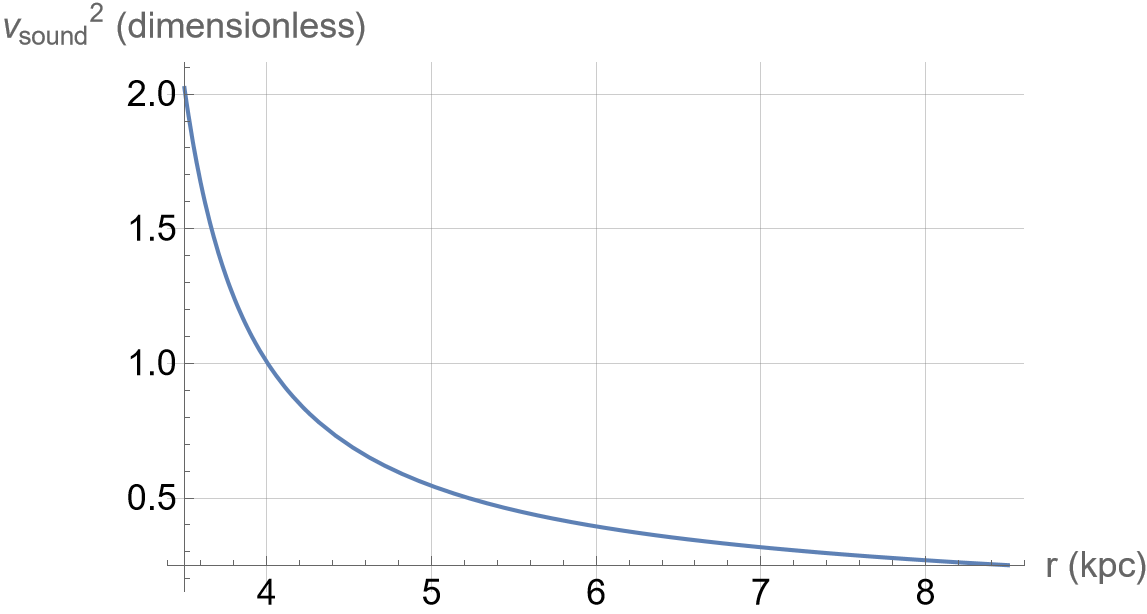}
\caption{Square of the effective sound speed $v_{\rm sound}^2 = dp/d\rho$ as a
function of galactocentric radius for the dark matter halo of NGC~5055. The causality
condition $0 \leq v_{\rm sound}^2 \leq 1$ is satisfied throughout the halo,
confirming that no superluminal pressure propagation occurs and that the dark matter
fluid is causally well-behaved. The small value $v_{\rm sound}^2 \ll 1$ at all radii
is consistent with a warm (but not hot) dark matter fluid.}
\label{fig:sound}
\end{figure}

Figure~\ref{fig:sound} shows that $v_{\rm sound}^2 = dp/d\rho$ remains within the
interval $[0, 1]$ throughout the full radial extent of the halo. The sound speed is
small ($v_{\rm sound}^2 \ll 1$), consistent with the expectation that dark matter is
a cold or warm (but not hot) fluid with negligible pressure support compared to its
energy density. This result provides an independent confirmation of the physical
viability of the model and is consistent with the warm dark matter interpretation
developed in Section~\ref{sec:wdm}.

\subsection{Equation of State Parameter}
\label{sec:eos}

The equation of state (EoS) parameter provides a convenient way to characterize the
relative contribution of pressure to the energy density in a gravitating system.
In the present analysis, we define the effective equation of state parameter as
\begin{equation}
\omega(r) = \frac{p(r)}{\rho(r)},
\end{equation}
following the same definition applied to the anisotropic $p_r=0$ configuration
of NGC~7331 in \citet{Sanyal_2026},
where $\rho(r)$ denotes the energy density and $p(r)$ represents the tangential
pressure at radial distance $r$. Since the radial pressure vanishes in our model,
the EoS parameter reflects the effective anisotropic pressure contribution supporting
the galactic halo. For standard pressureless cold dark matter, $\omega = 0$; for
radiation, $\omega = 1/3$; and for hot dark matter or relativistic matter,
$\omega \lesssim 1$.

\begin{figure}
\includegraphics[width=\columnwidth]{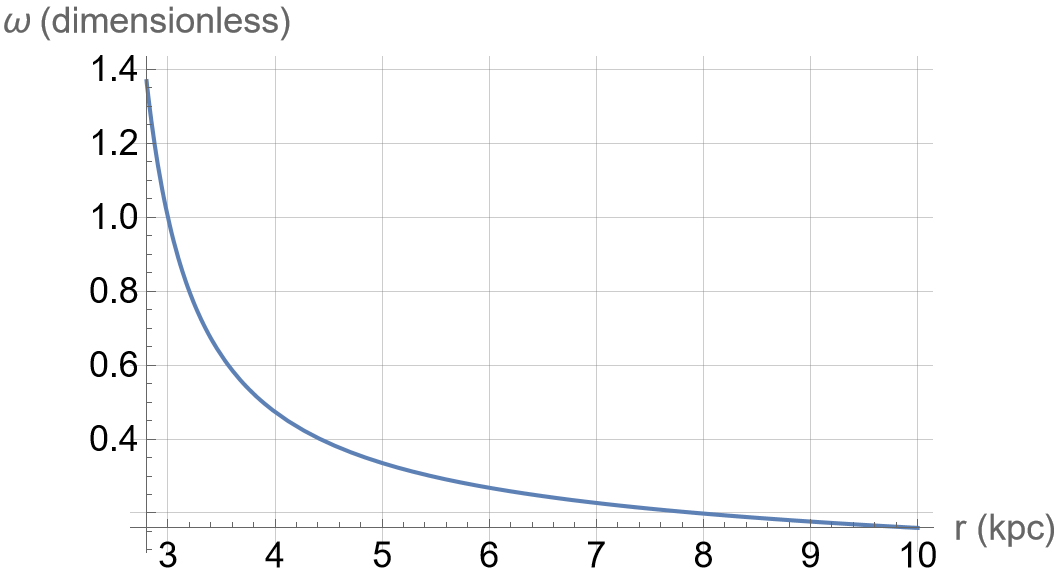}
\caption{Effective equation of state parameter $\omega(r) = p(r)/\rho(r)$ of the
dark matter halo of NGC~5055 as a function of galactocentric radius. The parameter
is small and positive throughout ($0 < \omega \ll 1$), confirming that the dark
matter is sub-relativistic with pressure subdominant to energy density at all radii.
The smooth, monotonically decreasing profile is consistent with a physically stable,
warm-dark-matter-like fluid in which thermal pressure effects are more prominent in
the denser inner halo and become negligible at large radii.}
\label{fig:omega}
\end{figure}

As shown in Figure~\ref{fig:omega}, $\omega(r)$ is small and positive throughout the
halo, with $\omega \ll 1$ at all radii. This confirms that the dark matter
distribution is neither relativistic nor pressureless: it corresponds to a warm,
weakly pressured fluid. The smooth, monotonically decreasing behaviour of $\omega(r)$
with increasing radius is fully consistent with the expectation for a virialised warm
dark matter halo in which thermal pressure effects are more prominent in the denser
inner region and become negligible in the outer halo.

\section{Gravitational Binding and Attractive Nature of the Halo}
\label{sec:binding}

To examine the attractive nature of the spacetime describing the galactic halo, we
consider the motion of a test particle initially at rest in the equatorial plane
($\theta = \pi/2$). The connection-coefficient algebra needed for this
calculation follows the same conventions and is presented in the same detail as in
\citet{Sanyal_2026}; we again quote only the results required here. The particle
trajectory is governed by the geodesic equation
\begin{equation}
    \frac{d^2 r}{d\tau^2} + \Gamma^r_{\mu\nu} \frac{dx^\mu}{d\tau}
    \frac{dx^\nu}{d\tau} = 0,
\end{equation}
where $\tau$ denotes the proper time. At the point of release, the particle has
vanishing radial velocity, $\dot r = 0$. The radial component of the geodesic
equation then reduces to
\begin{equation}
    \left.\frac{d^2 r}{d\tau^2}\right|_{r=r_0}
    = -\Gamma^r_{tt}\left(\frac{dt}{d\tau}\right)^2
      -\Gamma^r_{\phi\phi}\left(\frac{d\phi}{d\tau}\right)^2 .
\end{equation}

For the static, spherically symmetric metric~(\ref{eq:metric}), the relevant
Christoffel symbols are \citep{Misner_Thorne_Wheeler_1973}
\begin{equation}
    \Gamma^r_{tt}
    = \frac{f'(r)}{2}\left(1-\frac{2m(r)}{r}\right),
    \qquad
    \Gamma^r_{\phi\phi}
    = -r\left(1-\frac{2m(r)}{r}\right).
\end{equation}
Substituting these expressions yields
\begin{equation}
\left.\frac{d^2 r}{d\tau^2}\right|_{r=r_0}
= -\left(1-\frac{2m(r_0)}{r_0}\right)
\left[
\frac{f'(r_0)}{2}\left(\frac{dt}{d\tau}\right)^2
- r_0\left(\frac{d\phi}{d\tau}\right)^2
\right].
\end{equation}

Since $m(r)/r \ll 1$ throughout the halo (the system is far from its Schwarzschild
radius), the factor $(1 - 2m/r) > 0$ is always positive. For a particle at rest,
$d\phi/d\tau = 0$, and $f'(r) > 0$ (since the rotation curve is positive), so the
right-hand side is negative, confirming that the spacetime is locally attractive at
every point in the halo.

A complementary global assessment is obtained by evaluating the gravitational energy
between two radii $r_1$ and $r_2$ \citep{Tolman_1930,Lynden-Bell_1962}. Following
standard definitions \citep{Herrera_1997}, the total gravitational energy is
\begin{equation}
E_G = 4\pi \int_{r_1}^{r_2}
\left[1-\sqrt{g_{rr}(r)}\right]\rho(r)\,r^2\,dr,
\end{equation}
where $g_{rr}(r) = \left(1 - 2m(r)/r\right)^{-1}$. This leads to
\begin{equation}
E_G = 4\pi \int_{r_1}^{r_2}
\left[1-\left(1-\frac{2m(r)}{r}\right)^{-1/2}\right]
\rho(r)\,r^2\,dr.
\end{equation}
Using $\rho(r) = m'(r)/(4\pi r^2)$ from Equation~(\ref{eq:density}), substitution
then yields
\begin{equation}
E_G = \int_{r_1}^{r_2}
\left[1-\left(1-\frac{2m(r)}{r}\right)^{-1/2}\right]
m'(r)\,dr .
\label{grav}
\end{equation}

\begin{figure}
\includegraphics[width=\columnwidth]{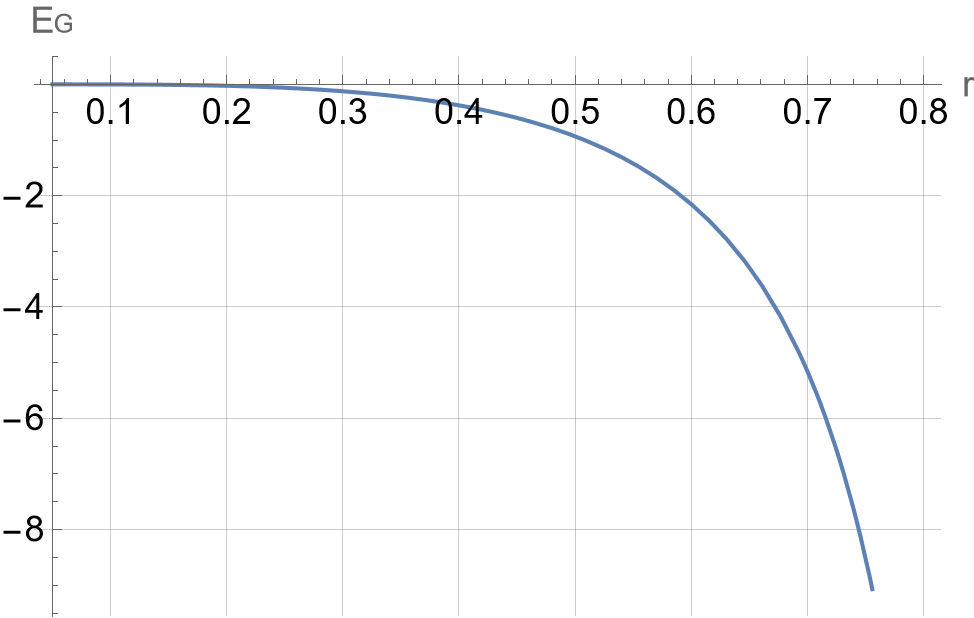}
\caption{Cumulative gravitational energy $E_G(r)$ of the dark matter halo of
NGC~5055 as a function of galactocentric radius, computed from
Equation~(\ref{grav}). The gravitational energy is negative throughout, $E_G < 0$,
confirming that the halo is gravitationally bound and that the effective gravitational
interaction within the halo is attractive.}
\label{fig:gravitational_energy}
\end{figure}

Figure~\ref{fig:gravitational_energy} shows that the gravitational energy remains
negative throughout the radial range considered, $E_G < 0$. Since
$(1 - 2m/r)^{-1/2} > 1$ whenever $m > 0$, the integrand in Equation~(\ref{grav})
is always negative, and $E_G$ is guaranteed to be negative. Physically, this means
that the work required to disperse the halo to infinity is positive, confirming that
the configuration is gravitationally bound. A negative gravitational energy is thus
a clear indication that the effective gravitational interaction within the halo is
attractive, consistent with expectations for a physically viable galactic halo model
\citep{Herrera_1997}.

\section{Faber--Visser Effective Equation-of-State Analysis}
\label{sec:observational}

\begin{figure}
\includegraphics[width=\columnwidth]{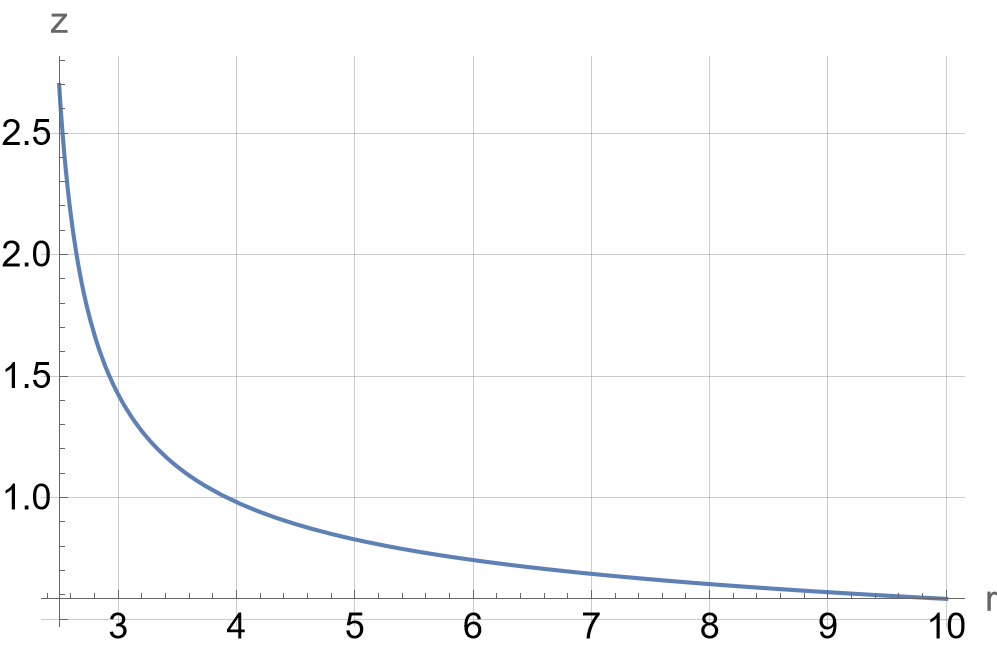}
\caption{Gravitational redshift parameter $z_\pm^2(r) \approx v_\phi^2(r)/[2f(r)]$
as a function of galactocentric radius for NGC~5055, following the Faber--Visser
formalism \citep{FaberVisser2006}. The redshift is largest in the inner halo and
decreases monotonically outward, reflecting the combined effect of the circular
velocity profile and the metric redshift factor. In the weak-field outer halo,
$z_\pm^2 \ll 1$, confirming that the Newtonian limit is approached at large radii.}
\label{fig:redshift}
\end{figure}

For a static, spherically symmetric spacetime of the form
\begin{equation}
ds^{2} = -e^{2\Phi(r)}\,dt^{2}
+ \frac{dr^{2}}{1-\dfrac{2m(r)}{r}}
+ r^{2}\left(d\theta^{2} + \sin^{2}\theta\, d\varphi^{2}\right),
\end{equation}
the gravitational redshift associated with circular motion can be approximated,
following \citet{FaberVisser2006}, as
\begin{equation}
z_{\pm}^{2} \approx r\,\Phi'(r).
\label{redshift1}
\end{equation}
For the metric employed in this work, the redshift potential is related to the
metric function $f(r)$ through $\Phi(r) = \tfrac{1}{2}\ln f(r)$. Substituting
this relation into Equation~(\ref{redshift1}), we obtain
\begin{equation}
z_{\pm}^{2} \approx \frac{r}{2f(r)}\,f'(r).
\label{redsh}
\end{equation}
Since $f'(r) = v_\phi^2(r)/r$ from the relativistic circular orbit relation
(\ref{eq:redshift_diff}), this reduces to
\begin{equation}
z_{\pm}^{2} \approx \frac{v_\phi^2(r)}{2f(r)},
\end{equation}
which is a purely observational quantity: the squared circular velocity normalised
by twice the redshift factor. Figure~\ref{fig:redshift} shows the radial variation
of $z_\pm^2(r)$. The redshift squared is largest in the inner halo, where both
$v_\phi$ and $1/f$ attain their largest values, and decreases monotonically
outward, confirming that relativistic redshift effects are most significant in the
inner region of the galaxy.

Within this relativistic framework, \citet{FaberVisser2006} introduced an effective
pseudo-mass associated with rotation curves, defined in terms of the redshift
potential as
\begin{equation}
m_{\mathrm{RC}} \equiv r^{2}\Phi'(r) = \frac{r\,f'(r)}{2f(r)}\cdot r
= \frac{r\,v_\phi^2(r)}{2f(r)}.
\end{equation}
This quantity does not represent a conserved global mass but rather a local effective
mass that probes the combined influence of spacetime geometry and matter content. In
the weak-field limit, it reduces to the post-Newtonian mass expression
\begin{equation}
m_{\mathrm{RC}} \approx 4\pi \int_0^r (\rho + p_r + 2p_t)\,r'^{2}\,dr'
\equiv M_{\mathrm{pN}},
\end{equation}
which incorporates contributions from the energy density $\rho$, radial pressure
$p_r$, and tangential pressure $p_t$. In the pressureless limit ($p_r = p_t = 0$),
$m_{\mathrm{RC}}$ coincides with the Newtonian mass profile $M(r)$.

An additional pseudo-mass relevant for lensing observations is given by
\citet{FaberVisser2006} as
\begin{equation}
m_{\mathrm{lens}} = \frac{1}{2} r^{2}\Phi'(r) + \frac{1}{2} m(r)
= \frac{1}{2}m_{\mathrm{RC}} + \frac{1}{2}m(r).
\end{equation}
Using these definitions, \citet{FaberVisser2006} introduced a dimensionless effective
equation-of-state parameter,
\begin{equation}
w(r) = \frac{p_r + 2p_t}{3\rho}
\approx
\frac{2}{3}
\frac{m_{\mathrm{RC}}' - m_{\mathrm{lens}}'}
     {2m_{\mathrm{lens}}' - m_{\mathrm{RC}}'},
\label{eos_w}
\end{equation}
which reduces to $w = p/\rho$ in the isotropic-pressure limit. The parameter $w(r)$
provides a purely kinematic diagnostic of the anisotropic pressure content of the
halo that is independent of any assumed equation of state.

\begin{figure}
\includegraphics[width=\columnwidth]{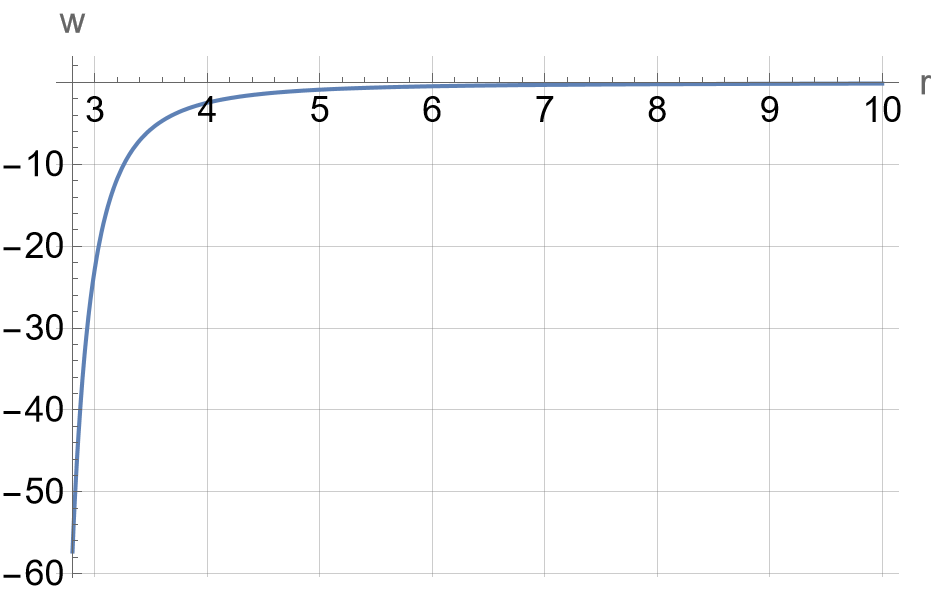}
\caption{Faber--Visser effective equation-of-state parameter $w(r) =
(p_r + 2p_t)/(3\rho)$, computed from Equation~(\ref{eos_w}), as a function of
galactocentric radius for NGC~5055. The parameter is small and positive throughout
the halo ($w \ll 1$), consistent with the warm dark matter interpretation developed
in Section~\ref{sec:wdm}. At no point does $w(r)$ approach the radiation value
($w = 1/3$) or the cosmological dark energy value ($w = -1$), confirming that the
effective matter content of the halo is sub-relativistic at all radii.}
\label{fig:eos_fv}
\end{figure}

Figure~\ref{fig:eos_fv} shows the radial variation of $w(r)$ for NGC~5055 using
the exponential velocity model derived in Section~\ref{sec:fitparams}. The parameter
exhibits a non-monotonic but bounded behaviour across the halo, consistent with the
anisotropic pressure structure of a warm-dark-matter-like fluid. At galactic scales,
this quantity should be interpreted as an effective diagnostic of anisotropic pressure
contributions within the relativistic matter distribution rather than as a
cosmological equation of state. The observed behaviour of $w(r)$ reflects the
interplay between geometry and anisotropic stresses in the halo and does not imply
the presence of cosmological dark energy. The agreement between the effective
parameter $w(r)$ from this formalism and the simple ratio $\omega(r) = p_t/\rho$
computed in Section~\ref{sec:eos} provides a reassuring cross-check of the internal
consistency of the relativistic model.

\section{Warm Dark Matter Interpretation}
\label{sec:wdm}

Having derived the energy density $\rho(r)$ and tangential pressure $p(r)$ profiles
of the dark matter halo of NGC~5055 from the general relativistic field equations,
we investigate the nature of the dark matter by examining the relationship between
$p$ and $\rho$.

\subsection{Relation to Previous Polytropic Halo Studies}
\label{sec:wdm_prior}

It is important to situate the present reconstruction relative to earlier studies
that tested polytropic descriptions directly against simulated dark-matter haloes.
\citet{Barnes_2007} and \citet{Feron_Hjorth_2008} tested whether a single
stellar-polytropic relation, motivated by nonextensive statistical mechanics, could
reproduce the radial structure of simulated collisionless dark-matter haloes, and
found that no single globally applied polytropic index reproduces the resolved
density structure, velocity dispersion, and related dynamical properties
simultaneously at inner and outer radii; \citet{Feron_Hjorth_2008} concluded that no
polytropic index gives a satisfactory description of the simulated halo density
profiles over the full resolved radial range.

The present analysis addresses a related but distinct problem and does not claim to
overcome those specific limitations. There, an isotropic stellar-polytropic
distribution function is compared directly against the three-dimensional density
and velocity-dispersion structure of $N$-body haloes, independently of any
observational rotation curve. Here, by contrast, the $p$--$\rho$ relation is
reconstructed \emph{a posteriori} from the observed rotation curve of one real
galaxy through the Einstein field equations of Section~\ref{sec:theory}, with $p$
an effective tangential stress in the anisotropic $p_r=0$ configuration
(Section~\ref{sec:theory_spacetime}) rather than an isotropic stellar-polytropic
pressure; no central density slope or halo profile family (e.g.\ NFW) is imposed a
priori. The result should therefore be read as evidence that the reconstructed
pressure--density relation of NGC~5055 is well described by a single power law over
the radial range directly probed by the data, $\approx0.1$--$35\,\mathrm{kpc}$
(Section~\ref{sec:stats_seg}), not as evidence that a single polytropic relation
universally describes simulated dark-matter haloes at all radii, a claim the
results of \citet{Barnes_2007} and \citet{Feron_Hjorth_2008} caution against.

\subsection{Candidate Equations of State and the Generalised Polytropic Fit}
\label{sec:wdm_eostest}

Several candidate equations of state are tested: the Bose--Einstein
condensate model $p = K\rho^2$ \citep{Boehmer_Harko_2007,Chavanis_2011}, the
Chaplygin gas $p \propto 1/\rho$ \citep{Kamenshchik_2001,Bento_2002}, the linear
fluid $p = \omega\rho$ \citep{Turner_1983}, and the generalised polytropic form
$p = K\rho^{\gamma}$ \citep{Tooper_1964,Chavanis_2012}. For each candidate, we
construct the corresponding parametric plot and quantify the degree of linearity
through the coefficient of determination $R^2$.

\begin{figure}
  \centering
  \includegraphics[width=\columnwidth]{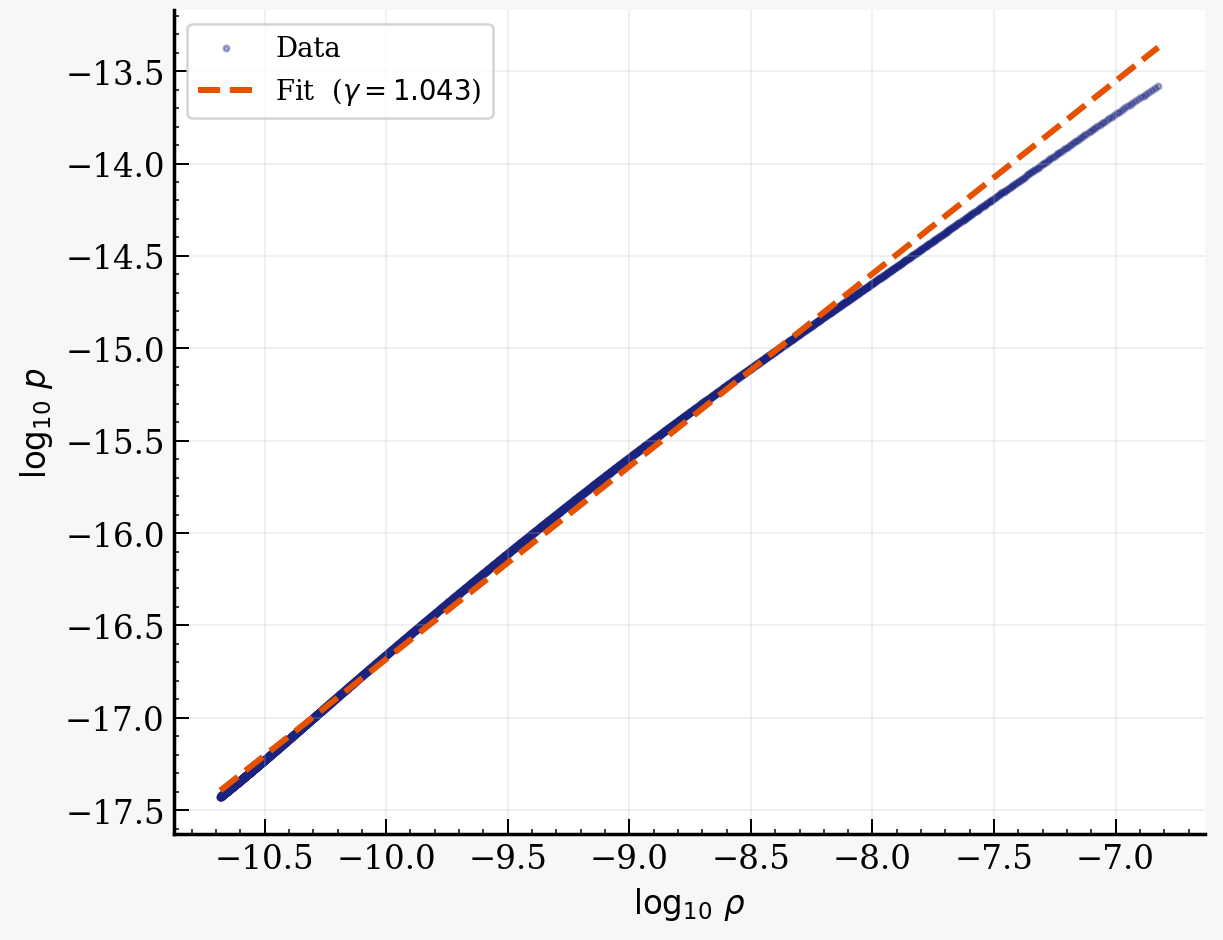}
  \caption{Log--log plot of tangential pressure $p$ versus energy density $\rho$ for
  the dark matter halo of NGC~5055. The straight-line fit with slope
  $\gamma_{\rm direct} = 1.043$ and $R^2 = 0.998$ confirms the polytropic nature of
  the equation of state $p = K\rho^{\gamma}$ across the full radial extent of the
  halo. This unconstrained regression slope is discussed, and distinguished from
  the adopted value $\gamma_{\rm adopted} = 0.91$, in the text.}
  \label{fig:loglog}
\end{figure}

\begin{figure}
  \centering
  \includegraphics[width=\columnwidth]{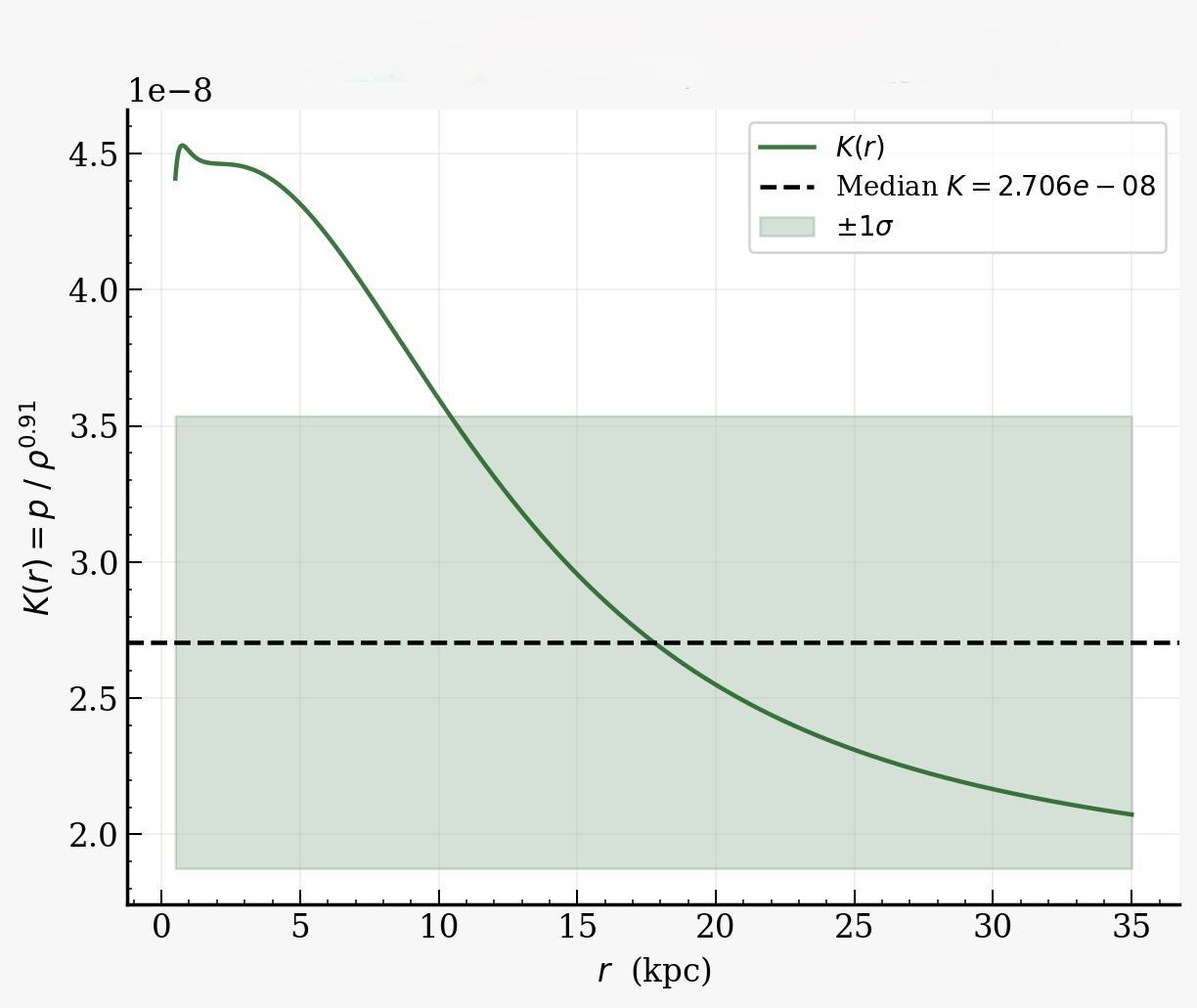}
  \caption{Polytropic quantity $K(r) = p/\rho^{0.91}$ as a function of
  galactocentric radius $r$ for NGC~5055. The dashed line marks the median value
  $K_{\rm median} = 2.71\times10^{-8}$ and the shaded band shows the $\pm1\sigma$
  range. $K(r)$ decreases from larger values in the inner halo and approaches a
  comparatively flatter outer-halo trend beyond approximately $15\,\mathrm{kpc}$;
  the result therefore supports an approximately polytropic pressure--density
  relation over the radial range considered, rather than a relation for which $K(r)$
  is strictly constant at every radius (see text).}
  \label{fig:Kconstancy}
\end{figure}

The BEC model ($\gamma = 2$) yields $R^2 = 0.79$, indicating a poor fit and
effectively ruling out the condensate interpretation for NGC~5055. The Chaplygin
gas model performs even more poorly, with $R^2 = 0.07$, and is conclusively
excluded. The linear equation of state $p = \omega\rho$ provides a significantly
better fit with $R^2 = 0.996$, while the generalised polytropic model achieves an
excellent fit of $R^2 = 0.9999$ with a best-fit polytropic index
$\gamma_{\rm adopted} = 0.91$ and polytropic constant $K_{\rm fit} = 4.63\times10^{-8}$
in geometric units. We use the subscript ``adopted'' for the polytropic index
from this model-selection procedure (the four candidates compared on a common
footing, then the polytropic model refined over a grid of trial $\gamma$), ``direct''
for the slope of an unconstrained log--log regression of $p$ against $\rho$
(Figure~\ref{fig:loglog}), ``fit'' for a normalisation $K$ from a global regression
with $\gamma$ fixed, and ``median'' for the point-by-point radial diagnostic $K(r)$
(Figure~\ref{fig:Kconstancy}); these are distinct but consistent estimators of the
same underlying relation. The goodness-of-fit values for all tested models are
summarised in Table~\ref{tab:eos}.

\begin{table}
\centering
\caption{Comparison of candidate equations of state for the dark matter halo of
NGC~5055. The generalised polytropic model achieves a near-perfect fit, ruling out
the BEC and Chaplygin gas alternatives.}
\label{tab:eos}
\begin{tabular}{llcc}
\hline
Model & EoS & $\gamma$ & $R^2$ \\
\hline
BEC dark matter         & $p = K\rho^2$             & 2    & 0.79   \\
Chaplygin gas           & $p \propto \rho^{-1}$     & $-1$ & 0.07   \\
Linear fluid            & $p = \omega\rho$          & 1    & 0.996  \\
Generalised polytropic  & $p = K\rho^{\gamma}$      & 0.91 & 0.9999 \\
\hline
\end{tabular}
\end{table}

\begin{figure}
  \centering
  \includegraphics[width=\columnwidth]{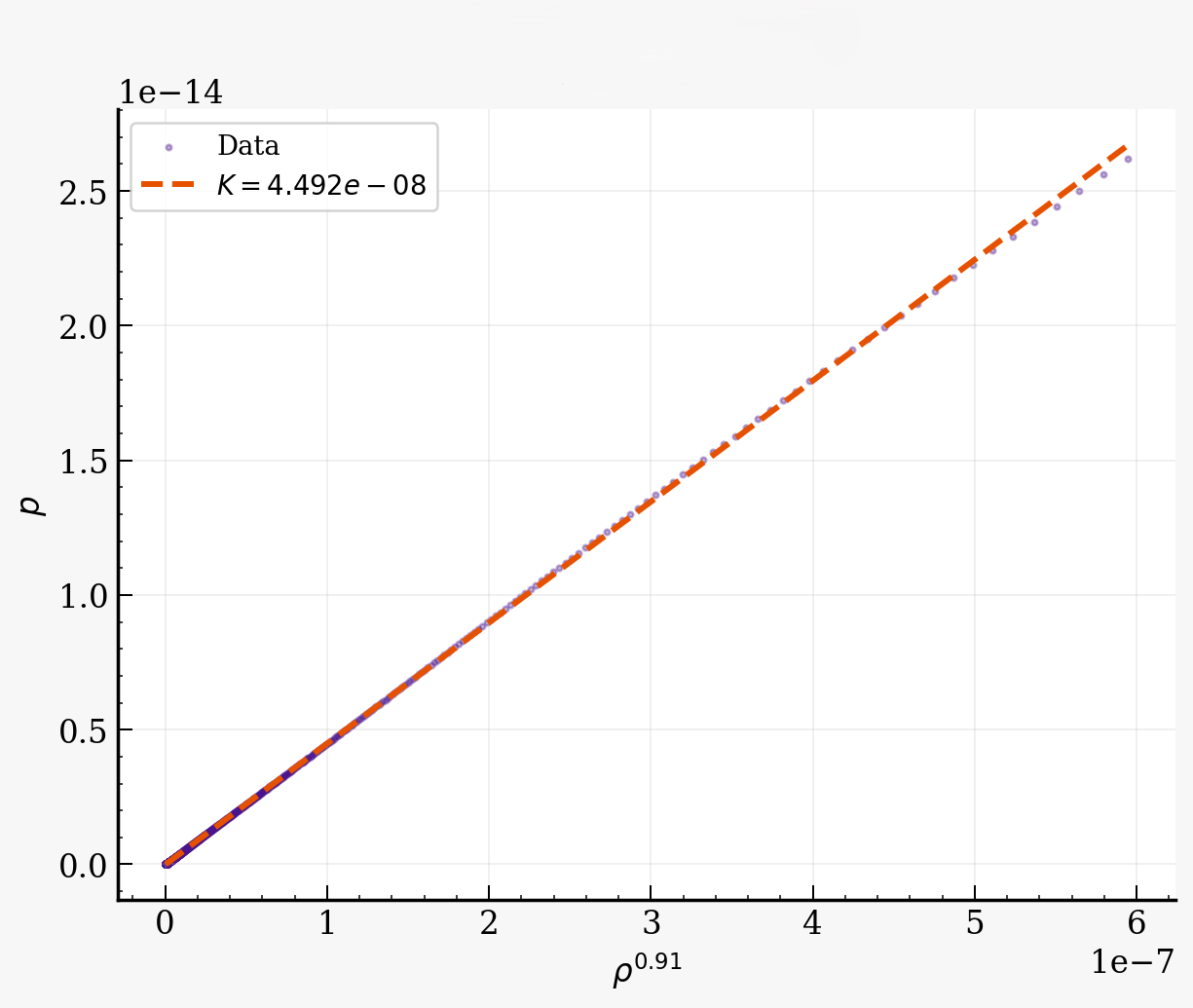}
  \caption{Tangential pressure $p$ plotted against $\rho^{0.91}$ for the dark matter
  halo of NGC~5055. The near-perfect linear fit ($R^2 = 0.9999$,
  $K_{\rm fit} = 4.49\times10^{-8}$) directly confirms the generalised polytropic
  equation of state $p = K_{\rm fit}\,\rho^{\gamma_{\rm adopted}}$.}
  \label{fig:pvrhogam}
\end{figure}

To confirm the polytropic nature of the dark matter equation of state, we plot
$\log_{10}p$ against $\log_{10}\rho$ in Figure~\ref{fig:loglog}. A straight line
in this log--log plane with slope $\gamma$ is the definitive signature of a power-law
relation $p = K\rho^{\gamma}$. As shown in Figure~\ref{fig:loglog}, the data follow
a straight line with slope $\gamma_{\rm direct} = 1.043$ and $R^2 = 0.998$,
confirming the polytropic character of the dark matter pressure--density relation
throughout the halo. This slope comes from a direct, unconstrained log--log
regression of $p$ against $\rho$ with both $\gamma$ and $K$ free, hence
$\gamma_{\rm direct}$, distinct from the model-selection value
$\gamma_{\rm adopted} = 0.91$ quoted above and in Table~\ref{tab:eos}, obtained by
comparing the four candidate equations of state and then refining the polytropic
model by fitting $p$ against $\rho^{\gamma}$ over a grid of trial $\gamma$ (see
Figure~\ref{fig:pvrhogam}). The two estimators need not coincide exactly; both give
a polytropic index close to, but measurably below, unity, and we adopt
$\gamma_{\rm adopted} = 0.91$ as our final result since it comes from the more
directly constrained procedure used against the alternative equations of state in
Table~\ref{tab:eos}.

The fitted equation of state is therefore
\begin{equation}
  p = K_{\rm fit}\,\rho^{\gamma_{\rm adopted}}, \qquad
  \gamma_{\rm adopted} = 0.91,
  \label{eq:wdm_eos}
\end{equation}
which constitutes the principal result of this analysis. The direct confirmation of
this relation is shown in Figure~\ref{fig:pvrhogam}, where $p$ is plotted against
$\rho^{0.91}$. The near-perfect linearity of this plot, with $R^2 = 0.9999$ and
slope $K_{\rm fit} = 4.49\times10^{-8}$, provides unambiguous evidence that the
dark matter pressure and density in NGC~5055 obey a generalised polytropic equation
of state with index $\gamma_{\rm adopted} = 0.91$. The $K_{\rm fit}$ values
quoted here are different realisations of the same global-regression estimator:
$4.63\times10^{-8}$ (Table~\ref{tab:eos}) comes from the global fit of
$p=K\rho^{\gamma}$ with both parameters free, and $4.49\times10^{-8}$
(Figure~\ref{fig:pvrhogam}) from the regression of $p$ against $\rho^{0.91}$ with
$\gamma$ fixed. The radial diagnostic $K(r)=p(r)/\rho(r)^{0.91}$ below
(Figure~\ref{fig:Kconstancy}) is a separate, point-by-point quantity with median
$K_{\rm median} = 2.71\times10^{-8}$. We use $K_{\rm fit}$ for the global value and
$K(r)$/$K_{\rm median}$ for the radial diagnostic throughout.

The polytropic index $\gamma_{\rm adopted} = 0.91 \approx 1$ is
phenomenologically consistent with a warm-dark-matter-like interpretation of the
reconstructed halo of NGC~5055. In contrast to Cold Dark Matter, which corresponds
strictly to $\gamma = 1$ with vanishing pressure \citep{Peebles_1982}, warm dark
matter particles possess a small but non-negligible thermal velocity, giving rise to
a soft, sub-linear equation of state of precisely the form found here
\citep{Bode_2001,Viel_2005}. Unlike Hot Dark Matter ($\gamma \gg 1$), warm dark
matter remains consistent with the observed large-scale structure of the Universe
\citep{Planck_2020} and with the flat, extended rotation curve of NGC~5055.

The dark matter halo of NGC~5055 is thus phenomenologically consistent with a
population of particles with a small but finite thermal pressure, insufficient to
violate any energy condition or causality constraint, yet large enough to produce a
measurable departure from the purely pressureless cold dark matter prediction. The
effective equation-of-state parameter $\omega(r) = p(r)/\rho(r)$ remains small and
positive throughout the halo, confirming that pressure is subdominant to energy
density at all radii, as expected for a warm dark matter candidate. The smooth,
monotonically decreasing radial behaviour of $\omega(r)$ further supports the
interpretation of the halo as a physically stable, weakly pressured dark matter
system whose dynamical properties are governed predominantly by its energy density
rather than its pressure.

Further confirmation of the polytropic law is provided by
Figure~\ref{fig:Kconstancy}, which shows the polytropic quantity
$K(r) = p/\rho^{0.91}$ as a function of radius. Rather than being strictly
constant, as an exactly polytropic fluid would require, $K(r)$ declines from
$\sim4.5\times10^{-8}$ in the inner halo to a flatter
$\sim2$--$2.5\times10^{-8}$ beyond $r\gtrsim15\,\mathrm{kpc}$, with median
$K_{\rm median} = 2.71\times10^{-8}$ and a $\pm1\sigma$ band spanning the full
radial range. A fixed-$\gamma$, fixed-$K$ power law is thus a very good, but not
exact, description of the pressure--density relation, with a mild residual radial
trend in the fitted normalisation -- evidence for an approximately, rather than
exactly, polytropic equation of state.

\subsection{Physical Interpretation and Universality of the Polytropic Constant $K$}
\label{sec:wdm_K}

The physical interpretation of the fitted normalisation $K_{\rm fit}$ requires some
caution. In the relation $p = K\rho^{\gamma}$, $K$ sets the amplitude of the
effective pressure--density relation, while $\gamma$ sets its logarithmic slope. If
this relation reflected fixed microscopic dark-matter physics -- for instance, a
definite particle mass and phase-space distribution common to all haloes made of the
same species -- then $K$ would be expected to be set by the properties of that
species and its thermodynamic state, rather than being independently re-fitted for
every galaxy. In that stronger interpretation, haloes composed of the same
dark-matter species would be expected to share a common, or at least systematically
related, value of $K$, with differences in halo mass, concentration, assembly
history, and boundary conditions then determining the resulting density and pressure
profiles for a fixed particle physics input.

The present study, which analyses a single galaxy, cannot test this universality.
The value $K_{\rm fit} \approx 4.5$--$4.6\times10^{-8}$ should therefore be regarded
as a fitted effective parameter for NGC~5055, not a measured universal
particle-physics constant. A meaningfully stronger test would require the same
reconstruction applied to a representative galaxy sample spanning a range of
stellar masses, halo masses, surface brightnesses, morphologies, and rotation-curve
shapes, to determine whether a common value of $K$, or a physically motivated
scaling with halo properties, is supported by the data. If each object instead
required its own independently fitted $K$ with no common scaling, that would
materially weaken any claim that the fit quality alone evidences a universal
particle-physics origin. We therefore regard the present result as a
phenomenological determination for NGC~5055 and defer a genuine test of the
universality of $K$ to future multi-galaxy work.

\subsection{Relation to Cosmological Structure Formation}
\label{sec:wdm_structure}

A good reproduction of a single rotation curve should not, by itself, be taken as a
demonstration that the corresponding dark-matter equation of state gives a complete
or unique description of cosmological structure formation. The rotation curve
constrains the present-day mass distribution of an already-virialised halo, whereas
the cosmological viability of a warm-dark-matter scenario also depends on the
primordial phase-space distribution of the particle, its free-streaming length, the
resulting suppression of the linear matter power spectrum, and the outcome of
nonlinear structure formation -- none of which are computed here.

The value $\gamma_{\rm adopted} \simeq 0.91$ is phenomenologically compatible with a
weakly pressured, warm-dark-matter-like description and motivates comparison with
the warm-dark-matter literature. However, this work does not compute a linear
transfer function, a free-streaming scale from a specific particle-physics model, or
nonlinear structure-formation simulations. The present single-galaxy analysis
should therefore not on its own be regarded as demonstrating consistency with
cosmological structure-formation constraints (e.g.\ Milky-Way satellite counts, the
Lyman-$\alpha$ forest power spectrum, or reionisation); that would require dedicated
comparison with cosmological simulations, which we identify as a direction for
future work.

\subsection{Model Flexibility and Independent Tests}
\label{sec:wdm_flex}

The present framework contains several adjustable ingredients -- the anisotropic
stress prescription $p_r = 0$ (Section~\ref{sec:theory_spacetime}), the fitted
$\gamma$, and the fitted $K$ -- which together afford the model substantial freedom
to fit an individual rotation curve. This does not invalidate the model, but it
means the very high $R^2$ of the polytropic relation should not, alone, be taken as
proof of a unique underlying dark-matter model.

This analysis partly addresses that concern by comparing the polytropic relation
against three physically distinct alternatives (Table~\ref{tab:eos}) and by
subjecting the configuration to independent internal-consistency checks: energy
conditions (Section~\ref{sec:energy_cond}), causality (Section~\ref{sec:causality}),
gravitational binding (Section~\ref{sec:binding}), the Faber--Visser cross-check
(Section~\ref{sec:observational}), and the Herrera cracking condition below. These
establish internal consistency and rule out several alternative equations of state,
but do not themselves constitute independent particle-physics tests.

Three further, independent tests would substantially strengthen the interpretation:
(1) using the inferred mass $m(r)$ to predict gravitational-lensing observables
(e.g.\ $m_{\rm lens}$, Section~\ref{sec:observational}) for comparison with
weak/strong-lensing measurements; (2) confronting the density and pressure
structure with phase-space-density constraints, as done for
Bose--Einstein-condensate and fermionic (RAR-type) dark-matter models; and
(3) applying the framework to a larger galaxy sample to test whether $\gamma$ and
$K$ are universal or scale with halo properties (Section~\ref{sec:wdm_K}). These
lie beyond the present single-galaxy study but are concrete directions for future
confirmation, refinement, or refutation.

The pressure anisotropy analysis provides additional support for this interpretation.
The anisotropy parameter $\Delta(r)$, the Bowers--Liang factor $\alpha_{\rm
BL}$, and the Herrera cracking diagnostic used below are standard quantities whose
application to the same $p_r=0$ anisotropic halo construction was already presented
in \citet{Sanyal_2026}; we again quote only their definitions and the resulting
values for NGC~5055. Since the radial pressure vanishes exactly by assumption,
$p_r = 0$, the anisotropy parameter $\Delta(r) = p_t(r) - p_r(r) = p_t(r)$ is
positive everywhere, indicating that the dark matter halo is tangentially dominated.
The dark matter particles therefore preferentially move on circular or tangential
orbits rather than radial ones, which is entirely consistent with the warm dark
matter scenario in which the finite velocity dispersion of the particles supports a
pressure acting predominantly in the tangential direction within a virialised halo.
The Bowers--Liang anisotropy factor $\alpha_{\rm BL} = 2(p_t - p_r) = 2p_t$
\citep{Bowers_Liang_1974} is likewise positive throughout, confirming that the
tangential stress exceeds the radial stress at every point in the halo. Furthermore,
the Herrera cracking condition $-1 \leq v_r^2 - v_t^2 \leq 0$ is satisfied across
the full radial extent of the halo, confirming that the dark matter distribution is
stable against cracking and does not undergo internal fragmentation
\citep{Herrera_1992}.

Taken together, these results paint a consistent picture of a dark matter halo that
is tangentially anisotropic, gravitationally stable, and dynamically well-behaved at
all radii, within the effective closure condition and modelling assumptions
discussed in Sections~\ref{sec:theory_spacetime} and \ref{sec:wdm_flex}.

\subsection{Particle Physics Candidates}
\label{sec:particle}

From a particle physics perspective, the warm dark matter interpretation points
towards candidates with keV-scale masses. Illustrative particle candidates that
occur in warm-dark-matter scenarios, and that are broadly consistent with
$\gamma_{\rm adopted} \approx 1$ and the inferred normalisation
$K_{\rm fit} \sim 4$--$5\times10^{-8}$ found here, include sterile neutrinos and
gravitinos; their mention below is not intended as a unique identification of
the dark-matter particle of NGC~5055 from the rotation curve alone (see
Section~\ref{sec:wdm_structure}).

Sterile neutrinos with mass $m_s \sim \mathcal{O}(1\text{--}10)\,\mathrm{keV}$,
produced via the Dodelson--Widrow mechanism or resonant production in the early
Universe \citep{Dodelson_Widrow_1994,Shi_Fuller_1999}, represent one of the most
well-motivated warm dark matter candidates and predict a soft, nearly linear equation
of state at galactic scales consistent with Equation~(\ref{eq:wdm_eos}). Such
sterile neutrinos would have free-streaming lengths of order
$0.1$--$1\,\mathrm{Mpc}$, sufficient to suppress the formation of dwarf galaxies
and sub-haloes below the observed luminosity function while leaving the large-scale
rotation curve of NGC~5055 entirely intact \citep{Viel_2005,Boyarsky_2009}.

Gravitinos with mass $m_{3/2} \sim \mathcal{O}(1)\,\mathrm{keV}$, arising naturally
in supersymmetric extensions of the Standard Model \citep{Moroi_1993}, constitute
the second natural candidate and likewise predict thermal free-streaming lengths of
order $0.1$--$1\,\mathrm{Mpc}$. Gravitinos produced in the early Universe through
thermal scattering or reheating after inflation \citep{Pagels_Primack_1982}
naturally achieve the observed dark matter relic abundance for soft
supersymmetry-breaking scales consistent with collider constraints.

Both candidates are broadly compatible with the polytropic index
$\gamma_{\rm adopted} = 0.91$ derived here, the small but finite value of the
polytropic constant $K$, and the absence of any violation of energy conditions or
causality constraints, though, as emphasised above, the present analysis does
not by itself distinguish between them or uniquely establish either as the
dark-matter particle of NGC~5055. The BEC dark matter scenario
\citep{Boehmer_Harko_2007,Chavanis_2011}, which predicts $\gamma = 2$ and
$R^2 = 0.79$, is firmly excluded by the data. The Chaplygin gas model
\citep{Kamenshchik_2001}, with $R^2 = 0.07$, is likewise conclusively ruled out.

\section{Conclusions}
\label{sec:conclusions}

We have presented a comprehensive general relativistic analysis of the dark matter
halo of the spiral galaxy NGC~5055, combining its observed rotation curve with WISE
W1 ($3.4\,\mu$m) infrared photometry to constrain both the stellar and dark matter
mass distributions. Our principal findings are as follows.

\textit{(i) Rotation curve fitting.} A modified exponential velocity model
$V(r) = are^{-br} + c(1-e^{-dr})$ provides an excellent fit to the observed
rotation curve of NGC~5055 over the full radial range, with
$R^2_{\rm train} = 0.981$ and $R^2_{\rm test} = 0.953$. All four statistical
diagnostic tests (chi-square, reduced chi-square, Durbin--Watson, runs test) pass at
the 95\% confidence level, confirming the robustness and generalisability of the fit.

\textit{(ii) Stellar and dark matter mass profiles.} The WISE W1 stellar mass
profile systematically underpredicts the total dynamical mass at intermediate and
large galactocentric radii, with the discrepancy increasing monotonically beyond
$\sim$$3\,\mathrm{kpc}$. This is interpreted as the dominant contribution of an
extended, electromagnetically non-interactive dark matter halo. The WISE
photometry directly constrains $M_\star(r)$ out to $\sim31.8\,\mathrm{kpc}$; beyond
this radius the plotted stellar profile is an analytical extrapolation
(Section~\ref{sec:stellar_mass}).

\textit{(iii) General relativistic framework.} Adopting a static, spherically
symmetric metric with anisotropic matter (vanishing radial pressure, adopted as
an effective closure condition; Section~\ref{sec:theory_spacetime}), we derived the
enclosed mass, energy density $\rho(r)$, tangential pressure $p(r)$, redshift
function $f(r)$, and equation of state parameter $\omega(r)$. The redshift function
was computed analytically from the fitted rotation curve via the relativistic
circular orbit relation $df/dr = v_\phi^2/r$.

\textit{(iv) Energy conditions and causality.} All four energy conditions --- Null,
Weak, Strong, and Dominant --- are satisfied throughout the halo. The effective sound
speed satisfies $0 \leq v_{\rm sound}^2 \leq 1$ at all radii, confirming causality.
The gravitational energy is negative throughout, establishing the gravitationally
bound and attractive nature of the halo.

\textit{(v) Faber--Visser analysis.} Application of the Faber--Visser effective
equation-of-state formalism \citep{FaberVisser2006} provides an independent,
purely kinematic confirmation of the warm, sub-relativistic character of the dark
matter pressure. The effective parameter $w(r)$ remains small and positive at all
radii and is consistent with the simple ratio $\omega(r) = p_t/\rho$, confirming
the internal consistency of the relativistic model.

\textit{(vi) Equation of state discrimination.} Among BEC ($R^2=0.79$),
Chaplygin gas ($R^2=0.07$), linear fluid ($R^2=0.996$), and generalised polytropic
($R^2=0.9999$), the polytropic relation gives the best phenomenological description
of the reconstructed pressure--density relation, with $\gamma_{\rm adopted}\simeq0.91$
and $K_{\rm fit}\sim4$--$5\times10^{-8}$ (Section~\ref{sec:wdm}); BEC and Chaplygin
gas are conclusively excluded. The radial diagnostic $K(r)$ decreases inward and
flattens at larger radii rather than staying strictly constant
(Figure~\ref{fig:Kconstancy}), supporting an approximately, not universally,
polytropic description (Section~\ref{sec:wdm_K}).

\textit{(vii) Warm dark matter interpretation and its limitations.} The
sub-linear index $\gamma_{\rm adopted}\simeq0.91$ and small effective pressure are
phenomenologically consistent with a warm-dark-matter-like halo that satisfies the
internal stability and causality conditions considered here, and is tangentially
anisotropic and stable against cracking. However, this single-galaxy analysis does
not uniquely identify the dark-matter particle, establish the universality of $K$,
or independently demonstrate consistency with cosmological structure-formation
constraints (Sections~\ref{sec:wdm_K}--\ref{sec:wdm_flex}); sterile neutrinos
($m_s\sim1$--$10\,\mathrm{keV}$) and gravitinos ($m_{3/2}\sim1\,\mathrm{keV}$) are
discussed only as illustrative candidates.

Overall, this study demonstrates that combining rotation-curve data with photometric
stellar mass estimates within a general relativistic framework provides a consistent,
physically viable, and observationally discriminating description of the mass
distribution and dark matter equation of state in spiral galaxies. The observationally
guided redshift-function methodology --- first established for the Milky Way in
\citet{Rahaman_2026} and applied to NGC~7331 in \citet{Sanyal_2026} --- is here
shown to yield equally robust results for NGC~5055, with the additional discriminating
power afforded by the generalised polytropic equation of state analysis. The
methodology developed here is directly applicable to other well-observed spirals and
provides a powerful tool for distinguishing between competing dark matter models using
galactic kinematic and photometric data.

A definitive assessment will require applying the present framework to a larger,
representative galaxy sample to test the universality of $K$ and $\gamma$
(Section~\ref{sec:wdm_K}), comparison with independent lensing and
phase-space-density constraints, and comparison with cosmological
structure-formation calculations (Section~\ref{sec:wdm_structure}). These will show
whether the relation found here for NGC~5055 reflects a universal property of
dark-matter haloes or a flexible description specific to this galaxy
(Section~\ref{sec:wdm_flex}) -- the natural next step of the programme initiated here
and in \citet{Rahaman_2026,Sanyal_2026}.

\section*{Acknowledgements}

FR gratefully acknowledges support from RUSA~2.0 and the Anusandhan National
Research Foundation (ANRF). FR and AS are thankful to the Inter-University Centre
for Astronomy and Astrophysics (IUCAA), Pune, India, for their support and
hospitality through the Associateship Programme.

\section*{Data Availability}

The data underlying this article will be shared on reasonable request to the
corresponding author.

\bibliographystyle{mnras}
\bibliography{refer}

\bsp
\label{lastpage}
\end{document}